\documentclass[%
preprint,
amsmath,amssymb,
aps,showkeys
]{revtex4-2}
\usepackage{graphicx}
\usepackage{dcolumn}
\usepackage{bm}
\usepackage{amsmath}
\usepackage{amssymb}
\usepackage{textcomp}
\usepackage{graphicx}
\usepackage{caption}
\usepackage{subcaption}
\usepackage{array}
\usepackage{mathtools}
\usepackage{centernot}
\usepackage{slashed}
\usepackage{hyperref}
\usepackage{xcolor}
\usepackage{tikz-feynman}
\usepackage{mhchem}
\newcommand*{\reddot}{\tikz\fill[red,radius=3pt] circle;}
\newcommand*{\orangedot}{\tikz\fill[orange,radius=3pt] circle;}
\newcommand*{\greendot}{\tikz\fill[green,radius=3pt] circle;}
\newcommand*{\bluedot}{\tikz\fill[blue,radius=3pt] circle;}

\hypersetup{
	colorlinks=true,       
	linkcolor=blue,          
	citecolor=magenta,        
	filecolor=cyan,      
	urlcolor=magenta           
}

\begin{document}
\title{Neutrinoless double beta decay in a supersymmetric left-right model}
\author{Vivek Banerjee}
\email{vivek\_banerjee@nitrkl.ac.in}
\author{Sasmita Mishra}
\email{mishras@nitrkl.ac.in}
\affiliation{Department of Physics and Astronomy, National 
Institute of Technology Rourkela, Sundargarh, Odisha, India, 769008}
	
	
\begin{abstract}
Neutrinoless double beta ($0\nu\beta\beta$) decay, an important 
low-energy process, serves not only as a potential test of the 
Majorana nature of neutrinos, but also as a sensitive probe for 
new physics beyond the Standard Model. In this study, the supersymmetric left-right model is explored to investigate its impact on $0\nu\beta\beta$ decay. Although the process takes 
place at low energies as compared to the electroweak scale, it carries the potential to provide indirect
hints about the parity-breaking scale $\text{M}_R$. In this work, we formulate the decay amplitude using an effective field theory 
approach by separating long- and short-range contributions, each 
expressed in terms of dimensionless particle physics parameters 
and nuclear matrix elements. The analysis shows that the $\text{M}_R$
must lie above $1$ TeV, and future experiments may
push it beyond $4 - 5$ TeV region. Another important outcome of
this work is the role played by the tentative dark matter candidates, the 
lightest neutralino and sneutrino, which contribute significantly to the half-life of $0\nu\beta\beta$
decay. This suggests that if any supersymmetric particle is detected
in future experiments, dark matter candidates will gain a 
permanent position in these extensions of the Standard Model.
\end{abstract}
	
\keywords{sneutrino, neutralino, neutrinoless double beta decay, supersymmetric left-right model}
	
\maketitle
\newpage   
	
\section{Introduction}
\label{sec:intro}
Neutrinoless double beta  ($0\nu\beta\beta$) decay (Ref.\cite{Doi:1991xf} for a review), one of the most 
important experimental avenues in particle physics, serves as a tabletop
experiment to probe for lepton number violation. Since the Standard Model (SM)
conserves lepton number, the affirmative detection of $0\nu\beta\beta$ decay would guarantee the presence of physics beyond the SM (BSM).  This enhances its utility as a 
low-energy probe to test various BSM scenarios by offering phenomenological 
constraints. For example, the parameter space for R-parity-violating \cite{Babu:1995vh, Hirsch:1995cg} and R-parity-conserving \cite{Hirsch:1997dm} theories receives important constraints from their
contributions to $0\nu\beta\beta$ decay. Similarly, models including leptoquarks are also constrained \cite{Hirsch:1996qy}. Left-right symmetric models have likewise placed stringent bounds on their parameters \cite{Hirsch:1996qw, Chakrabortty:2012mh, BhupalDev:2014qbx, Borah:2016iqd, Barry:2013xxa, Banerjee:2023aro}. The well-known tree-level diagram that confirms the Majorana nature of neutrinos \cite{Klapdor-Kleingrothaus:2000tjb, Schechter:1981bd, Klapdor-Kleingrothaus:1999tta, Baudis:1999xd} 
represents the simplest interpretation of $0\nu\beta\beta$ decay, arising from the fusion of two standard beta decay processes. 
Besides the presence of the $V-A$ vertices, there may be effective vertices 
involving scalar, pseudoscalar, and tensor currents that can induce the decay.
Also, the effective couplings can give rise to short-range contributions
that are not mediated by light neutrino exchange. Whereas the characteristic
scale of momentum transfer in $0\nu\beta\beta$ decay is $\simeq 100$ MeV,
the scale of the BSM that can induce the decay can lie above it.
In an effective field theory approach, the authors of Ref.\cite{Dvali:2023snt} show that the scale of new physics,
$\Lambda\simeq 3$ TeV, requiring that the neutrino Majorana mass 
contribution to $0\nu\beta\beta$ decay is subdominant as compared to 
that of the BSM physics. 
In this work, we study the new contributions that can induce the $0\nu\beta\beta$ decay in the supersymmetric version of left-right 
symmetric model \cite{Mohapatra:1974gc, Senjanovic:1975rk}, 
referred to as the supersymmetric left-right model (SUSYLRM). In an effective field theory approach, we study
$0\nu\beta\beta$ decay by taking the parity-breaking scale 
$M_R$ as the scale of new physics.

A notable class of such BSM models is the left-right 
symmetric model (LRSM) \cite{Mohapatra:1974gc, Senjanovic:1975rk},
governed by
the gauge group $\mathcal{G}_{\text{LRSM}} = SU(3)_C \times SU(2)_L
\times SU(2)_R \times U(1)_{B-L}$. A key feature of LRSM is that, in
the limit $v_R \rightarrow \infty$, it smoothly converges to the SM gauge group, where $v_R$ is directly related to the characteristic parity breaking scale, the scale of new physics \cite{Deshpande:1990ip}.  LRSMs naturally accommodate finite neutrino masses, 
which are absent in the SM, through the incorporation of both type I \cite{Minkowski:1977sc,Ramond:1979py,Gell-Mann:1979vob,  Sawada:1979dis,Levy:1980ws} and type II seesaw \cite{Magg:1980ut, Lazarides:1980nt, Mohapatra:1980yp,  Schechter:1980gr} mechanisms. 
Due to its connection with Grand Unified Theory (GUT) \cite{Deshpande:1990ip},
LRSMs also offer a platform to explore high-energy
phenomena. While it restores right-handed isospin symmetry, the LRSM
still falls short in addressing key open questions in particle physics, 
such as flavor anomalies, the nature of dark matter, and the hierarchy problem. 
The key advantage of SUSYLRM over LRSM is the inclusion
of the lightest supersymmetric particle (LSP) as a dark matter candidate. Its benefits
can be summarized as follows: (1) it facilitates neutrino mass
generation, (2) it provides a natural explanation for parity violation,
(3) it addresses strong and weak CP problems without invoking axions \cite{PhysRevD.54.5835, PhysRevLett.76.3486, Mohapatra:1995xd},
and (4) it explains the absence of excessive CP violation. The supersymmetric version
of the LRSM can
be constructed by expanding the scalar field content of the LRSM in two distinct ways: by introducing additional scalar
triplets \cite{Aulakh:1997fq, Aulakh:1997ba, Hirsch:2015fvq} or by including an extra singlet \cite{Babu:2008ep, PhysRevD.83.073007, Frank:2014kma} to achieve parity breaking
while preserving $R$-parity.
For simplicity, we adopt the latter approach. The number of new
parameters in SUSYLRM is significantly larger than in LRSM, presenting
challenges for phenomenological studies. Therefore, we employ standard
benchmark points that are consistent with multiple observational
constraints. 

In the SUSYLRM framework, the complete Lagrangian for 
$0\nu\beta\beta$ decay can be formulated in terms of effective 
dimensionless particle physics parameters corresponding to both 
long-range (two intermediate vertices connected via a neutrino
propagator) and short-range interactions.
Each effective vertex term involves a factor of $1/ \text{M}_{\text{R}}$,
which is crucial for extracting information about the parity-breaking scale. 
In this study, each effective vertex includes multiple
vertices connecting off-shell SUSYLRM fields
with on-shell SM particles. Due to mixing among the charginos,
neutralinos, and sneutrinos, each diagram contains multiple vertex
coefficients corresponding to the associated interaction terms. All the
loops with vertices where SUSYLRM meets SM can be broken into simple
sub-interaction vertices. The possible sub-interactions and their respective
vertex coefficients are provided in the appendix (\ref{app:sub_interactions}). 
The particle physics parameters encapsulate both the vertex coefficients and
the propagator contributions, where the latter are handled using the
Passarino-Veltman reduction scheme. It is expected that the contribution from the
SUSYLRM diagrams are suppressed as compared to the standard diagrams involving
light neutrinos. This is because they arise from higher orders in perturbation
theory and suffer loop suppression due to the presence of heavy sparticles
as intermediate states.
	
The paper is organized as follows. In Section (\ref{sec:SUSYLR}), we discuss
the structure and features of the SUSYLRM
model. In section (\ref{app:formalism}), we discuss the general formalism 
of calculating the half-life of $0\nu\beta\beta$ decay in an effective field theory 
approach. Section (\ref{sec:onbb}) elaborates on the general Lagrangian 
for $0\nu\beta\beta$ decay in the SUSYLRM context, along with the 
relevant interactions. Numerical estimates and results are presented 
in the section.(\ref{sec:result}), followed by the concluding remarks in 
Section.(\ref{sec:discussion}).
		
\section{Supersymmetric left-right model}
\label{sec:SUSYLR}
SUSYLRM, the dedicated supersymmetric extension of LRSM,
is based on the gauge group $SU(3)_C \times
SU(2)_L \times SU(2)_R \times U(1)_{B-L}$. 
Since $ B-L$ is a conserved quantity in this framework, the theory
inherently preserves $R$-parity, defined as $R_P = (-1)^{3(B-L) + 2s}$.
This built-in feature of $R$-parity distinguishes SUSYLRM from MSSM, enhancing its theoretical appeal.
Its superfield contents in the quark and lepton sectors, by suppressing the generation indices, are
\begin{equation}
Q_L\left(3,2,1,\frac{1}{3}\right) = \begin{pmatrix}
u_L \\ d_L 
\end{pmatrix}, \quad 
Q_R\left(3^*,1,2,-\frac{1}{3}\right) = \begin{pmatrix}
d_R \\ -u_R
\end{pmatrix},
\label{eqn:quarks_SUSYLR}
\end{equation}
\begin{equation}
L_L\left(1,2,1,-1\right) = \begin{pmatrix}
\nu_L \\ l_L
\end{pmatrix}, \quad 
L_R\left(1,1,2,1\right) = \begin{pmatrix}
l_R \\ -\nu_R
\end{pmatrix}.
\label{eqn:leptons_SUSYLR}
\end{equation}
In the Higgs sector, the bidoublet is
doubled to have a nonvanishing Cabbibo-Kobayashi-
Maskawa mixing matrix, whereas the triplets are doubled to
have anomaly cancelation.
The Higgs superfields are represented as
\begin{equation}
\Delta_{1L}(1,3,1,-2) = \begin{pmatrix}
\frac{\delta_{1L}^+}{\sqrt{2}} & \delta_{1L}^{++}  \\
\delta_{1L}^0 & -\frac{\delta_{1L}^+}{\sqrt{2}}
\end{pmatrix}, \quad
\Delta_{2L}(1,3,1,2) = \begin{pmatrix}
\frac{\delta_{2L}^+}{\sqrt{2}} & \delta_{2L}^{++}  \\
\delta_{2L}^0 & -\frac{\delta_{2L}^+}{\sqrt{2}}
\end{pmatrix},
\label{eqn:LH_higgs_SUSYLR}
\end{equation} 
and their right-handed counterparts are\begin{equation}
\Delta_{1R}(1,1,3,-2) = \begin{pmatrix}
\frac{\delta_{1R}^+}{\sqrt{2}} & \delta_{1R}^{++}  \\
\delta_{1R}^0 & -\frac{\delta_{1R}^+}{\sqrt{2}}
\end{pmatrix}, \quad
\Delta_{2R}(1,1,3,2) = \begin{pmatrix}
\frac{\delta_{2R}^+}{\sqrt{2}} & \delta_{2R}^{++}  \\
\delta_{2R}^0 & -\frac{\delta_{2R}^+}{\sqrt{2}}
\end{pmatrix}.
\label{eqn:RH_higgs_SUSYLR}
\end{equation}
The bidoublets are 
\begin{equation}
\Phi_1 (1,2,2^*,0) = \begin{pmatrix}
\phi_1^+ & \phi_1^{0'} \\
\phi_1^0 & \phi_1^-
\end{pmatrix}, \quad 
\Phi_2 (1,2,2^*,0) = \begin{pmatrix}
\phi_2^+ & \phi_2^0 \\
\phi_2^{0'} & \phi_2^-
\end{pmatrix}.
\label{eqn:bidoublets_SUSYLR}
\end{equation}
With the introduction of right-handed Higgs triplets, the
breaking of $SU(2)_R$ can be achieved, leading to the generation of small neutrino masses
via the seesaw mechanism. Within this framework, it was shown in Ref. \cite{Kuchimanchi:1993jg, Kuchimanchi:1995vk} that spontaneous parity breaking cannot be achieved at the renormalizable level without
breaking $R$-parity simultaneously.	In this case, the vacuum may favor a solution in which the
right-handed sneutrino acquires a non-zero vacuum expectation
value (VEV). 	
There are two main approaches to circumvent this issue: 
(i) \textbf{Introducing an additional singlet field ($S$)} : 
The singlet helps achieve a stable, $R$-parity-conserving 
vacuum once one-loop corrections are incorporated into the scalar
potential \cite{Babu:2008ep}.
(ii) \textbf{Including two additional Higgs triplets with $(B-L) = 0$}:
These fields can spontaneously break the left-right symmetry
while preserving $R$-parity at the tree level \cite{Aulakh:1998nn,
Aulakh:1997ba, Aulakh:1997fq}.
To maintain simplicity in the analytical framework and calculations,
we adopt the first approach. The additional singlet field is represented 
as,
\begin{equation}
S \sim (1,1,1,0),
\end{equation}
where the numbers in parentheses denote the transformation properties
under the gauge group $SU(3)_C \times SU(2)_L \times SU(2)_R \times 
U(1)_{B-L}$. 	
The superpotential in this SUSYLRM is given by
\begin{equation}
\begin{split}
W  = &  \hspace{5pt}Y_Q^1 Q_L^T \sigma_2 \Phi_1 \sigma_2 Q_R + Y_Q^2 Q_L^T  \sigma_2 \Phi_2 \sigma_2 Q_R +  Y_L^1 L_L^T \sigma_2
\Phi_1 \sigma_2 L_R + Y_L^2 L_L^T \sigma_2 \Phi_2 \sigma_2 L_R \\& + i(Y_L^3 L_L^T  \sigma_2 \Delta_{2L} L_L + 
Y_L^4 L_R^T \sigma_2 \Delta_{1R} L_R) + S[\lambda_L \text{Tr}(\Delta_{1L}
\Delta_{2L}) + \lambda_R \text{Tr}(\Delta_{1R}\Delta_{2R}) \\&+ \lambda_3
\text{Tr}(\Phi_1^T \sigma_2 \Phi_2 \sigma_2) + \lambda_4 \text{Tr}
(\Phi_1^T \sigma_2 \Phi_1 \sigma_2) + \lambda_5 \text{Tr}(\Phi_2^T 
\sigma_2 \Phi_2 \sigma_2) + \lambda_S S^2 - \text{M}^2_R].
\end{split}
\label{eqn:superpotential}
\end{equation}
Here, $Y_Q^1$, $Y_Q^2$ are the Yukawa coupling matrices for quarks, and
$Y_L^1$, $Y_L^2$, $Y_L^3$, $Y_L^4$ are for leptons. $\lambda_L$, $\lambda_R$
$\lambda_3$, $\lambda_4$, $\lambda_5$ and $\lambda_S$ are the trilinear
couplings of the theory and the term $\text{M}_R$ refers to the parity breaking scale of SUSYLRM \cite{Babu:2008ep}. 
The superpotential shown in Eq.(\ref{eqn:superpotential}) remains invariant under the parity transformation. The fields transform under parity as,
\begin{equation}
\begin{split}
    & Q_L \rightarrow Q_R^*, \quad L_L \rightarrow L_R^*, \quad \Delta_{1L} \rightarrow \Delta_{1R}^*, \quad \Delta_{2L} \rightarrow \Delta_{2R}^*,
    \\& \phi_1 \rightarrow \phi_1^{\dagger}, \quad \phi_2 \rightarrow \phi_2^{\dagger}, \quad S \rightarrow S^*.
\end{split}
\end{equation}
The parity-invariant condition leads to the hermiticity of the Yukawa coupling matrices $Y_Q^1, Y_Q^2, Y_L^1$ and $Y_L^2$.  For Majorana-type Yukawa couplings, the relation $Y_L^3 = (Y_L^4)^*$ is achieved. The parity-invariant condition also assures the real value of $\lambda_3, \lambda_4, \lambda_5$ and $M_R^2$. The vacuum structure of the model is reflected in the VEVs of the neutral components of the Higgs fields. They are represented as,
\begin{equation}
\begin{split}
&\langle \Delta_{1R} \rangle = \begin{pmatrix}
0 & \frac{v_{1R}}{\sqrt{2}} \\
0 & 0
\end{pmatrix}, \quad \langle \Delta_{2R}\rangle = \begin{pmatrix}
0 & 0 \\
\frac{v_{2R}}{\sqrt{2}} & 0
\end{pmatrix}, \\&
\langle \Phi_1 \rangle = \begin{pmatrix}
0 & \frac{v'_1}{\sqrt{2}}e^{i\alpha_1} \\
\frac{v_1}{\sqrt{2}} & 0
\end{pmatrix}, \quad
\langle \Phi_2 \rangle= \begin{pmatrix}
0 & \frac{v_2}{\sqrt{2}} \\
\frac{v'_2}{\sqrt{2}}e^{i\alpha_2} & 0
\end{pmatrix}, \\&
\langle S \rangle = \frac{v_S}{\sqrt{2}}s^{i\alpha_S}.
\end{split}
\label{eqn:vevs_SUSYLR}
\end{equation}
The VEVs $v_{1R}, v_{2R},v_1, v_2$ and $v_S$ are taken as real and positive. From the CP-violating $W_L^{\pm}-W_R^{\pm}$ mixing and $K^0 - \bar{K}^0$ oscillation data and the constraint of $SU(2)_R$ gauge bosons, it demands large values for the VEVs originating in the right sector and imposes two useful conditions regarding the VEVs, which are,
\begin{equation}
    v_S, v_{1R},v_{2R} \gg v_2,v_1, \quad \text{and} \quad
    v'_1 = v'_2 = \alpha_1 = \alpha_2 =\alpha_S \simeq 0.
\end{equation}
And within the supersymmetric limit, after the parity breaking, the F- and D-flatness conditions give rise to three relations,
\begin{equation}
    \lambda_R v_{1R}v_{2R} = \text{M}^2_R, \quad \lvert v_{1R} \lvert = \lvert v_{2R} \lvert, \quad \langle S \rangle=0.
\end{equation}
The soft SUSYLRM Lagrangian is given by,
\begin{equation}
		\begin{split}
			\mathcal{L}_{\text{soft}} = & -\frac{1}{2}[M_1\tilde{B}\tilde{B} 
			+ M_{2L}\tilde{W}^a_L\tilde{W}_{La} +  M_{2R}\tilde{W}^a_R
			\tilde{W}_{Ra}+ M_3\tilde{g}^a\tilde{g}_a + \text{h.c.} ] \\&
			-m_{\Delta 1L}^2 \text{Tr}(\Delta_{1L}^{\dagger}\Delta_{1L}) 
			-m_{\Delta 2L}^2 \text{Tr}(\Delta_{2L}^{\dagger}\Delta_{2L})
			-m_{\Delta 1R}^2 \text{Tr}(\Delta_{1R}^{\dagger}\Delta_{1R}) \\&
			-m_{\Delta 2R}^2 \text{Tr}(\Delta_{2R}^{\dagger}\Delta_{2R})
			-m_{\Phi 1}^2 \text{Tr}(\Phi_1^{\dagger}\Phi_1)
			-m_{\Phi 2}^2 \text{Tr}(\Phi_2^{\dagger}\Phi_2) -m_S^2 \lvert S
			\lvert^2 \\& +m^2_{\tilde{Q}_L}\tilde{Q}_L^{\dagger}\tilde{Q}_L
			-m^2_{\tilde{Q}_R}\tilde{Q}_R^{\dagger}\tilde{Q}_R
			-m^2_{\tilde{L}_L}\tilde{L}_L^{\dagger}\tilde{L}_L
			-m^2_{\tilde{L}_R}\tilde{L}_R^{\dagger}\tilde{L}_R \\& -S[(T_L 
			\text{Tr}(\Delta_{1L}\Delta_{2L}) + T_R \text{Tr}(\Delta_{1R}
			\Delta_{2R}) + T_3 \text{Tr}(\Phi_1^T \tau_2 \Phi_2 \tau_2) \\& 
			+ T_4 \text{Tr}(\Phi_1^T \tau_2 \Phi_1 \tau_2) + T_5 \text{Tr}
			(\Phi_2^T\tau_2 \Phi_2 \tau_2) + T_S S^2 + \zeta_s] + \text{h.c.}
			\\& + [ T_Q^1 \tilde{Q}_L^T \phi_1 \tilde{Q}_R + T_Q^2 
			\tilde{Q}_L^T \phi_2\tilde{Q}_R + T_L^1 \tilde{L}_L^T \phi_1 
			\tilde{L}_R + T_L^2 \tilde{L}_L^T \phi_2 \tilde{L}_R  \\& 
			+ T_L^3 \tilde{L}_L^T \Delta_{2L} \tilde{L}_L +
			T_L^4 \tilde{L}_R^T \Delta_{1R} \tilde{L}_R + \text{h.c.}].
		\end{split}
		\label{eqn:soft_SUSYLR_lrgn}
	\end{equation}
The soft Lagrangian has a large number of new parameters.
Masses for all the gauginos ($M_1$, 
$M_{2L}$, $M_{2R}$, $M_3$ ), squarks ($m_{\tilde{Q}_L}$, $m_{\tilde{Q}_R}$),
sleptons ($m_{\tilde{L}_L}$, $m_{\tilde{L}_R}$), the scalar triplets 
($m_{\Delta 1L}$, $m_{\Delta 2L}$, $m_{\Delta 2L}$, $m_{\Delta 2R}$), 
bidoublets ($m_{\Phi_1}$,$m_{\Phi_2}$) and the corresponding couplings
($T_L$, $T_R$, $T_3$, $T_4$, $T_5$, $T_S$, $T_Q^1$, $T_Q^2$, $T_L^1$, $T_L^2$, $T_L^3$, $T_L^4$), offer a challenging 
situation for studying the phenomenological aspects.
It is useful to define or restrict the parameter
values using experimental and theoretical constraints. This led
to pinning down some benchmark points (BPs), which fix the parameters
in different sets of values following the experimental, cosmological, and
theoretical constraints. In Section (\ref{sec:result}), these BPs are 
discussed and are used in this study.   	
	
\subsection{Neutrino mass in SUSYLRM}
\label{sec:nu_mass_SUSYLR}
The seesaw mechanism, the obvious process of generation of a small neutrino mass,
appears in SUSYLRM on its own merit. After electroweak
symmetry-breaking, the part of the Lagrangian carrying the neutrino mass takes the form \cite{Frank:2002hk},
\begin{equation}
-2 \mathcal{L}_{\text{mass}} = \bar{\nu}_L^c \text{M}_{\nu}\nu_R +
\bar{\nu}_R^c \text{M}^*_{\nu}\nu_L^c.
\label{eqn:nu_mass_lagn_SUSYLR}
\end{equation} 
	
The $(6\times6)$ mass matrix $\text{M}_{\nu}$ for neutrinos is expressed 
as,
\begin{equation}
\text{M}_{\nu} = \begin{pmatrix}
0 & m_D \\ m_D^{\dagger} & M_N
\end{pmatrix}.
\label{eqn:nu_mass_matrix_SUSYLR}
\end{equation}
Here, the Dirac mass term is defined as $m_D = \frac{1}{\sqrt{2}}(Y_L^1 v_1 + Y_L^2
v'_2) \simeq \frac{1}{\sqrt{2}}Y_L^1 v_1$, where the heavy Majorana mass term is $M_N = \frac{1}{\sqrt{2}} Y^4_L v_{1R}$. 	
The $\langle \Delta_{1L/2L} \rangle = v_L = 0$ as the corresponding field
would acquire a small VEV only after electroweak symmetry breaking. Diagonalizing the mass matrix in Eq.(\ref{eqn:nu_mass_matrix_SUSYLR}) will provide
the seesaw masses for small neutrinos, as well as the masses for heavy neutrinos. In the seesaw limit, the light neutrino mass matrix is given by
\begin{equation}
m_{\nu} = -m_D \frac{1}{M_N} m_D^{\dagger}.
\end{equation} 
Now, after diagonalizing the mass
For the matrix $m_{\nu}$, one can obtain the mass of light neutrino mass states. The
mass eigenvalues have a direct contribution to the $0\nu\beta\beta$ decay, as shown in
section (\ref{sssec:1nd_category}). 

\section{General formalism of $0\nu\beta\beta$ decay}
\label{app:formalism}
The origin of neutrinoless double beta $0\nu\beta\beta$ decay can be traced back to the simplest weak interaction process: single beta decay. The nuclear physics underlying ordinary $\beta$ decay has been extensively studied and well established since early times. In this context, $\beta$ emitting isotopes are often represented on two distinct energy parabolas: 1. \textbf{Odd-odd nuclei}, which occupy higher energy states, 2. \textbf{Even-even nuclei}, which lie at comparatively lower energy states. These parabolas are instrumental in identifying nuclei that are candidates for beta-decay processes. Interestingly, the optimal nuclei for observing $0\nu\beta\beta$ decay are the same as those favorable for the two-neutrino double beta $2\nu\beta\beta$ decay. Conceptually, $0\nu\beta\beta$ decay can be visualized as arising from the merging of two successive single beta decay processes, but occurring within the same nucleus and without the emission of neutrinos. Some example of such nuclei are $\ce{^{76}_{32}\text{Ge}}, \ce{^{136}_{54}\text{Xe}}, \ce{^{128}_{52}\text{Te}}$, and $\ce{^{48}_{20}\text{Ca}}$ \cite{Zuber:2012fd}. The most general tree-level Lagrangian for $\beta$-decay in SM \cite{Pas:1999fc}, 
\begin{equation}
		\mathcal{L}_{\text{Weak}} = \frac{G_F}{\sqrt{2}} j^{\mu}_{V-A} 
		J^{\dagger}_{V-A,\mu}.
\label{eqn:sim_lg}
\end{equation}
Here, $j^{\mu}_{V-A}$ and $J_{V-A,\mu}$ refer to the left-handed leptonic and
hadronic currents, respectively, and $G_F$ is the Fermi coupling
constant. The forms of the currents are
\begin{equation}
j^{\mu}_{V-A} = \bar{e}\gamma^{\mu}(1 - \gamma_5)\nu, \quad J^{\mu}_{V-A} = \bar{u}\gamma^{\mu}(1 - \gamma_5)d.
\end{equation}
This Lagrangian is restricted within the SM itself. So, for introducing BSM contributions, the simplest
Lagrangian needs to be extended in a form \cite{Pas:1999fc},
	\begin{equation}
		\mathcal{L}_{\text{Weak}} = \frac{G_F}{\sqrt{2}}\left[ j^{\mu}_{V-A} 
		J^{\dagger}_{V-A,\mu} + \sum_{\alpha\beta}\eta_{\alpha\beta}
		j_{\alpha}J^{\dagger}_{\beta}\right].
		\label{eqn:sim_lg_2}
	\end{equation} 
In Eq.(\ref{eqn:sim_lg_2}), the second term is the BSM term, which
includes an effective coupling $\eta_{\alpha\beta}$, which depends on the leptonic and hadronic currents present at the effective BSM vertex. In the regime of the SM gauge group, the currents include only $V-A$ terms. But in BSM models, the currents
can be expressed, in general, as
\begin{equation}
		j_{\alpha} = \bar{e}\mathcal{O}_{\alpha}\nu, \quad J_{\beta} = 
		\bar{u}\mathcal{O}_{\beta}d,
\label{eqn:currents}
\end{equation}
where the operators $\mathcal{O}_{\alpha}$ and $\mathcal{O}_{\beta}$
can take the forms,
\begin{equation}
\mathcal{O}_{V \pm A} = \gamma^{\mu}(1 \pm \gamma_5), \quad 
\mathcal{O}_{S \pm P} = (1 \pm \gamma_5), \quad \mathcal{O}_{T_{R/L}}
= \frac{i}{2}[\gamma_{\mu},\gamma_{\nu}](1 \pm \gamma_5).
\label{eqn:operators}
\end{equation}
The effective coupling $\eta_{\alpha\beta}$ carries the
signature of the model in which the decay process is being studied.
In general, they include the lepton 
number violating parameters, so in this study, one of our objectives
is to find out the possible $\eta$ terms for 
$0\nu\beta\beta$ decay in the SUSYLRM.

\begin{figure}[h]
		\centering
		\begin{minipage}{0.3\textwidth}
			\centering
			\begin{subfigure}{\linewidth}
				\includegraphics[width=\linewidth]{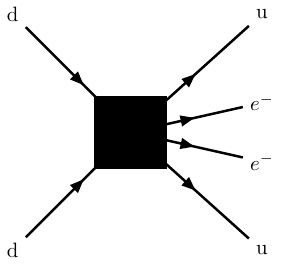}
				\caption{}
				\label{fig:a}
			\end{subfigure}
		\end{minipage}
		\hfill
		\begin{minipage}{0.6\textwidth}
			\centering
			\begin{tabular}{ccc}
				\begin{subfigure}{0.28\textwidth}
					\includegraphics[width=\linewidth]{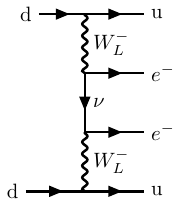}
					\caption{}
					\label{fig:b}
				\end{subfigure} &
				\begin{subfigure}{0.28\textwidth}
					\includegraphics[width=\linewidth]{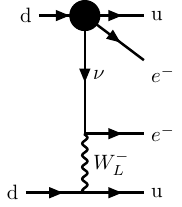}
					\caption{}
					\label{fig:c}
				\end{subfigure} &
				\begin{subfigure}{0.28\textwidth}
					\includegraphics[width=\linewidth]{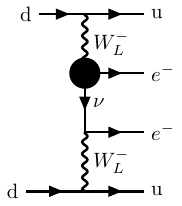}
					\caption{}
					\label{fig:d}
				\end{subfigure} \\
				\begin{subfigure}{0.28\textwidth}
					\includegraphics[width=\linewidth]{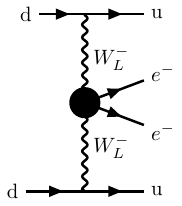}
					\caption{}
					\label{fig:e}
				\end{subfigure} &
				\begin{subfigure}{0.28\textwidth}
					\includegraphics[width=\linewidth]{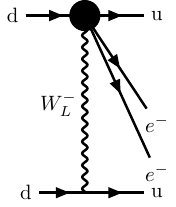}
					\caption{}
					\label{fig:f}
				\end{subfigure} &
				\begin{subfigure}{0.30\textwidth}
					\includegraphics[width=\linewidth]{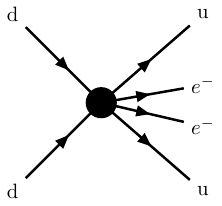}
					\caption{}
					\label{fig:g}
				\end{subfigure}
			\end{tabular}
		\end{minipage}
\caption{The possible classes of effective vertices for 
neutrinoless double beta decay connecting the SM to BSM 
theories. Figure (\ref{fig:a}) refers to the complete effective vertex diagram, which can be decomposed into the rest of the diagrams. On the right side, all three diagrams in the upper line
are long-range in nature, whereas the rest of the three diagrams below are short-range interactions.}
\label{fig:effective_vertices}
\end{figure}

Now, to study the $0\nu\beta\beta$ decay properties, one needs 
to club two tree level $\beta$ decay Lagrangians in a time
ordered manner. So the $0\nu\beta\beta$ beta decay amplitude is proportional to the term
\cite{Pas:1999fc},
\begin{equation}
	T(\mathcal{L}^1 \mathcal{L}^2) = \left(\frac{G_F}{\sqrt{2}}\right)^2
		T\left(\underbrace{j_{V-A} J^{\dagger}_{V-A} j_{V-A}  
			J^{\dagger}_{V-A}}_{\text{(pure SM term)}} + \underbrace{ 
			\eta_{\alpha\beta}j_{\alpha}  J^{\dagger}_{\beta}j_{V-A}
			J^{\dagger}_{V-A}}_{(\text{SM meets BSM})} + \underbrace{
			\eta_{\alpha\beta}\eta_{\delta\gamma}j_{\alpha} J^{\dagger}_{\beta}
			j_{\delta} J^{\dagger}_{\gamma}}_{(\text{pure BSM terms})}\right).
\label{eqn:time_ord_lgn}
\end{equation} 
In Eq.(\ref{eqn:time_ord_lgn}), the first term refers 
the basic diagram for $0\nu\beta\beta$ decay, which leads to the test of the
possible Majorana nature of the neutrino. The second term comes
from an effective vertex that will generate the possible decays of $0\nu\beta\beta$,
which connects the SM with BSM theories. And the last term is completely BSM in nature, and the presence of the product
of two $\eta$ terms makes this one insignificant in terms of 
contribution. So, using the first and second terms, a number 
of classes for the effective vertices can be constructed. 
They are shown in Fig.(\ref{fig:effective_vertices}).
		
In Fig.(\ref{fig:effective_vertices}), the diagrams shown from figures (\ref{fig:b}) to (\ref{fig:f}) have separate SM vertices along with effective BSM vertices, where in the diagram in Fig.(\ref{fig:a}), a single compact, effective vertex is observed, which suggests the general effective Lagrangian in Eq.(\ref{eqn:Lgn_0nbb_LRSYSY}). Following this general class of diagrams, the effective amplitude for $0\nu\beta\beta$ decay can be constructed, which will lead to the decay width for the process. So, the effective
amplitude for $0\nu\beta\beta$ decay takes the form (based
on the effective diagrams shown in Fig.(\ref{fig:effective_vertices})),
\begin{equation}
		\Lambda_{0\nu\beta\beta}^{eff} = \left<(A, Z+2),2e^- \left\lvert
		\mathcal{T} \text{exp}\left[i\int d^4x (\mathcal{L}_{0\nu\beta\beta}^{eff}) \right] \right\lvert (A,Z)\right>.
\label{eqn:0nbb_amplitude}
\end{equation}
In Eq.(\ref{eqn:0nbb_amplitude}), $\mathcal{L}_{0\nu\beta\beta}^{eff}$ is the effective Lagrangian, which one can construct depending on the model parameters. The currents, usually, are expressed in terms of 
quarks but the amplitude in Eq.(\ref{eqn:0nbb_amplitude})
written in terms of nucleons because the $0\nu\beta\beta$ 
decay is a nuclear process taking place at the level of nuclear 
transition. The initial
$(A,Z)$ and final $(A, Z+2)$ nuclear states are 
involved in the decay processes for which the information is captured in the nuclear matrix 
elements (NME) for a particular atom. The nonrelativistic impulse approach (NRIA) 
is one of the useful methods for the
calculation of NMEs. Using this amplitude, the decay rate for
$0\nu\beta\beta$ decay is calculated following the second 
order Fermi's Golden rule. The differential decay width is given by, 
\begin{equation}
d\Gamma_{0\nu\beta\beta}^{(0_i^+ \rightarrow 0_f^+)} = 2\pi 
		\sum_{\text{spin}} \lvert \Lambda_{0\nu\beta\beta}^{eff} 
		\lvert^2 \delta(E_1 + E_2 + E_f - E_i) 
		d\Omega_{e_1} d\Omega_{e_2}. 
		\label{eqn:decay_width}
\end{equation}
Here, $\Lambda_{0\nu\beta\beta}^{eff}$ is the amplitude shown for the decay of $0\nu\beta\beta$, and $E_1$ and $E_2$ are the energies of the emitted electrons. $E_i$ ($E_f$) is the energy of the initial state, $\lvert 0_i^+\rangle$ (final state, $\lvert 0_f^+ \rangle$) of the nucleus. The terms $d\Omega_{e1}$ and $d\Omega_{e2}$ refer to the infinitesimal phase space volumes of the emitted electrons. 

In this section, our objective was to provide a comprehensive overview of the fundamentals of $0\nu\beta\beta$ decay.
The excellent articles in \cite{Kotila:2012zza,Doi:1982dn,
Doi:1985dx, Muto:1989cd, Pantis:1992qe, Suhonen:1998ck} may be referred to to get a complete
idea of the process from the beginning with wave functions, following numerous
approximations and methods, and finally formulating the final expression
for decay width and half-life.
\section{$0\nu\beta\beta$ decay in SUSYLRM}
\label{sec:onbb}	
Neutrinoless double beta decay has the potential to detect not only the Majorana nature of
neutrinos but also other BSM physics \cite{Dvali:2023snt}.
The search for SUSY imprints with decay $0\nu\beta\beta$ has been carried out in reference \cite{Mohapatra:1986su, Feng:2002ev, Hirsch:1997dm}. In this study,
we will look for the SUSYLRM signatures in $0\nu\beta\beta$.  As discussed in the previous section, the term in Eq.(\ref{eqn:time_ord_lgn}) that connects SM to BSM is $\eta_{\alpha\beta}j_{\alpha}  J^{\dagger}_{\beta}j_{V-A}J^{\dagger}_{V-A}$. The contribution of this term provides diagrams with a single effective vertex, and one can write down the effective Lagrangian for $0\nu\beta\beta$ decay with the help of a new-physics scale, i.e., $\text{M}_R$, which indicates the breaking of SUSYLRM to the minimal supersymmetric standard model (MSSM).

As shown in Fig. (\ref{fig:effective_vertices}), the diagram in Fig.(\ref{fig:a}), where the effective vertex is represented as a solid square, refers to the effective diagram that can be decomposed into SM (Fig.
(\ref{fig:b})) and SUSYLRM (Fig.\ref{fig:c} to Fig.\ref{fig:g}). Correspondingly, the Lagrangian for this case is
\begin{equation}
    \mathcal{L}_{0\nu\beta\beta}^{eff}  = \mathcal{L}_{Wf\bar{f}} + \mathcal{L}_{0\nu\beta\beta}^{\text{SUSYLRM}}.
    \label{eqn:eff_lagrangian}
\end{equation}
The first term on the right-hand side comes from the SM term as shown in Eq.(\ref{eqn:time_ord_lgn}), corresponding to Fig.(\ref{fig:b}).
Now, following the diagrams Fig.(\ref{fig:c}) to Fig.(\ref{fig:g}), the corresponding effective Lagrangian terms in SUSYLRM can be written as \cite{Hirsch:1997dm},
\begin{equation}
\begin{split}
    \mathcal{L}_{0\nu\beta\beta}^{\text{SUSYLRM}}  & =
			  \frac{\eta_2}{\text{M}^{n_1}_R} J^{\mu} \bar{e} 
			(\mathcal{O}_1)_{\mu}\nu_L^c + \frac{\eta_3}{\text{M}^{n_2}_R} W^-_{\mu} \bar{e} 
			(\mathcal{O}_2)^{\mu} \nu_L^c  + 
			\frac{\eta_4}{\text{M}^{n_3}_R} W^-_{{\mu}}W^-_{\nu} \bar{e} 
			(\mathcal{O}_3)^{\mu\nu}e^c \\& + \frac{\eta_5}{\text{M}^{n_4}_R} 
			J^{\mu}W^-_{\mu} \bar{e} (\mathcal{O}_4)e^c + \frac{\eta_6}{\text{M}^{n_5}_R} J^{\mu}J^{\nu} \bar{e} (\mathcal{O}_5)_{\mu\nu}e^c.
\end{split}
\label{eq:susulr-eff}
\end{equation}

For simplicity of representation, the $\alpha$ and $\beta$ indices in the $\eta$ terms and the generation indices are suppressed in the Lagrangian. From the first to the sixth term in the right-hand-side of Eq.(\ref{eq:susulr-eff}), each term of the Lagrangian includes an effective vertex $\eta_i$, which is defined as $\eta_i$, $i$ = 2 to 6, and they represent the diagrams from Fig.(\ref{fig:c}) to Fig.(\ref{fig:g}) respectively. Each of the $\eta$ terms is written in such a way that it associates the fields directly connected to the effective vertices only (can be seen from diagrams in Fig.(\ref{fig:effective_vertices})), and the scale $\text{M}_R$ is used to make the mass dimension for each contribution of the effective Lagrangian equal to four ($d = 4$). So, one can explicitly determine the values of the indices ($n_1, n_2, n_3, n_4$ and $n_5$) on the powers of $1/\text{M}_R$ simply by following the criteria, $d = 4$. Considering the diagram in Fig.(\ref{fig:c}), and the corresponding effective Lagrangian term in Eq.(\ref{eqn:Lgn_0nbb_LRSYSY}), i.e. the second term in the right-hand side includes the $\eta_2$ as the dimensionless particle physics parameter, and the $J^{\mu}$ is the quark current associated with the effective vertex along with the electron and neutrino fields. Because the SM field contents have a total mass dimension of 6, they are regularized with the help of $\text{M}^2_R$ in the denominator, which implies that $n_1 = 2$ is chosen for the second term. The next terms are also written using the same formalism, and the values of the indices are determined as ($n_2=0, n_3=1, n_4=3, n_5=5$). So, the final expression for the Lagrangian in SUSYLRM becomes
\begin{equation}
	\begin{split}
			\mathcal{L}_{0\nu\beta\beta}^{\text{SUSYLRM}} & = 
              \frac{\eta_2}{\text{M}^2_R} J^{\mu} \bar{e} 
			(\mathcal{O}_1)_{\mu}\nu_L^c + \eta_3 W^-_{\mu} \bar{e} 
			(\mathcal{O}_2)^{\mu} \nu_L^c  + 
			\frac{\eta_4}{\text{M}_R} W^-_{{\mu}}W^-_{\nu} \bar{e} 
			(\mathcal{O}_3)^{\mu\nu}e^c \\& \hspace{15pt} + \frac{\eta_5}{\text{M}^3_R} 
			J^{\mu}W^-_{\mu} \bar{e} (\mathcal{O}_4)e^c + \frac{\eta_6}
			{\text{M}^5_R} J^{\mu}J^{\nu} \bar{e} (\mathcal{O}_5)_{\mu\nu}e^c.
		\end{split}
		\label{eqn:Lgn_0nbb_LRSYSY}
\end{equation}
This Lagrangian can also be separated into two parts: (1) long-range
\cite{Pas:1999fc} and (2) short-range \cite{Pas:2000vn}. Usually,
long-range interactions are defined as those in which a neutrino serves as a mediator between two vertices. The rest of the effective
vertices are short-range because of the very small lifetime of the W boson, whose propagator length is so small that it can be considered as a point-like interaction. So, in Fig.(\ref{fig:effective_vertices}), the first three are the long-range interactions, and the rest are short-range interactions. So, in our calculation, the long and short ranges
are treated accordingly to obtain the half-life formula.
	
\subsection{First category: Long-range interactions}
\label{ssec:long-range}
\label{sssec:1nd_category}
\begin{figure}[h]
		\includegraphics[height=4.4cm,width=4cm]{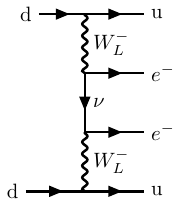}
		\caption{The simplest possible diagram for $0\nu\beta\beta$ decay.}
		\label{fig:1}
	\end{figure}

The first category of effective vertex arises from the first term $\mathcal{L}_{Wf\bar{f}}$
in Eq.(\ref{eqn:eff_lagrangian}). The diagram representing
this term is shown in Fig.(\ref{fig:1}).
This is the most fundamental tree-level diagram for $0\nu\beta\beta$ decay. The disappearing neutrinos in this process can be promising proof for the Majorana nature of neutrinos. Being the fundamental one, it includes no effective vertex, and in general, the straightforward formula of half-life for this case can be written as
\begin{eqnarray}
 \left[T_{0\nu\beta\beta}^{1/2}\right]^{-1} = G_{01} \lvert \mathcal{M}
\lvert^2 \lvert \eta \lvert^2 = G_{01} \lvert \mathcal{M} \lvert^2 \left\lvert\left(\frac{m_{eff}^{0\nu\beta\beta}}{m_e}\right)\right\lvert^2,   
\end{eqnarray}
where the effective electron neutrino mass has the form 
$m_{eff}^{0\nu\beta\beta} = \lvert\sum_{i=1}^{3}U_{ei}^2 m_i \lvert$.
Here, $U$ is the Pontecorvo-Maki-Nakagawa-Sakata (PMNS) 
neutrino mixing matrix. And, $m_i$ are the mass eigenvalues of the
neutrino mass matrix.
The $G_{01}$, $\mathcal{M}$ are the phase-space factor and nuclear
matrix element, respectively. This particular diagram has the popular reverse Y-shaped (famous as lobster plot) region as the parameter space when plotting the half-life with respect to the smallest neutrino mass in
the normal (inverted) hierarchical case. However, in this study, this contribution is not taken into account to highlight the contribution of SUSYLRM. Mainly, the diagrams where on-shell standard model fields
are connected with off-shell SUSYLRM fields, and are studied extensively. 
The rest of the categories are discussed below, and all of them follow
the above criteria.
	
\subsection{Second category: Long-range interactions}
\label{sssec:2nd_category}
\begin{figure}[h]
		\centering
		\begin{subfigure}[b]{0.4\textwidth}
			\centering
			\includegraphics[height=5cm,width=5cm]{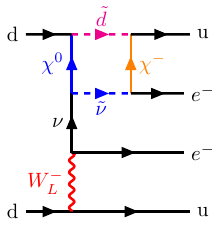}
			\caption{}
			\label{fig:2a}
		\end{subfigure}
		\hspace{35pt}
		\begin{subfigure}[b]{0.4\textwidth}
			\centering
			\includegraphics[height=5cm,width=5cm]{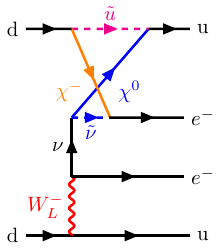}
			\caption{}
			\label{fig:2b}
		\end{subfigure}
		\caption{The two diagrams represent the possible SUSYLRM diagrams for the 
			effective diagram shown in Fig.(\ref{fig:c}).}
		\label{fig:2nd_category}
	\end{figure}

Now, coming to the intended study, following the first long-range 
effective vertex as shown in Fig.(\ref{fig:c}), which originates from the first term on the right side of $\mathcal{L}_{0\nu\beta\beta}^{\text{SUSYLRM}}$ in Eq.(\ref{eqn:Lgn_0nbb_LRSYSY}), can be expanded 
in two possible diagrams within SUSYLRM. In each diagram, the SM
gauge bosons and fermions are drawn in red and black, respectively.
For the SUSYLRM fields, charged sfermions and charginos are shown in pink and yellow, respectively.
The neutral fermions and neutralinos are colored
blue. For the second category, the possible $R$-parity-conserving diagrams are shown in Fig.(\ref{fig:2nd_category}).
	
The corresponding dimensionless effective coupling constants 
for Fig.(\ref{fig:2a}) and Fig.(\ref{fig:2b}) are,
\begin{equation}
		\begin{split}
			& \eta_{2a} = V_{ud} V_{(e^- - W_L^- - \nu)} V_{(\tilde{\nu}
				- \chi^0 - \nu)} V_{(e^- - \chi^- - \tilde{\nu})} V_{(u - \tilde{d}
				- \chi^-)} V_{(d - \tilde{d} - \chi^0)} \sum_{a=\tilde{d}, 
				\tilde{\nu}, \chi^0, \chi^-} \frac{x_a^2 \ln{x_a}}{\prod_{a\neq b}
				(x_b - x_a)}, 
			\\& \eta_{2b} = V_{ud}. V_{(e^- - W_L^- - \nu)}. V_{(\tilde{\nu} 
				- \chi^0 - \nu)} V_{(e^- - \chi^- - \tilde{\nu})} V_{(d - \tilde{u}
				- \chi^-)} V_{(u - \tilde{u} - \chi^0)} \sum_{a=\tilde{u}, 
				\tilde{\nu}, \chi^0,\chi^-} \frac{x_a^2 \ln{x_a}}{\prod_{a\neq b} 
				(x_b - x_a)}. 
		\end{split}
		\label{eqn:2nd_category}
	\end{equation}
   
The two diagrams in Fig.(\ref{fig:2nd_category}) consist of several vertices where three fields meet. Each of these vertices includes a vertex factor depending on the nature of the interactions. For example, in Fig.(\ref{fig:2a}), the number of such vertices is six, and they are expressed as the $V$-terms in the corresponding dimensionless particle physics parameter $\eta_{2a}$. All possible vertex factors and their forms are shown in the appendix(\ref{app:sub_interactions}). In the expressions of Eq.(\ref{eqn:2nd_category}), the dimensionless
particle physics parameters $\eta_{2a}$ and $\eta_{2b}$ are formed
by accumulating all the vertex factors present in each diagram and
multiplying it by the loop in terms of the propagators present in
it. Because each of the diagrams includes multiple propagators
in loop, the Passarino-Veltman reduction technique \cite{Passarino:1978jh}
is used to simplify the integrals arising from the propagators.
Here, each of the mass terms involved in the propagators is
scaled with respect to the mass of the SM W boson, and the term $x_a$ 
is expressed as $x_a = \frac{m_a^2}{m_W^2}$. Here, the total contribution from the second category is $\eta_2 = \eta_{2a} + \eta_{2b}$. 
	
\subsection{Third category: Long-range interactions}
\label{sssec:3rd_category}
\begin{figure}[h]
		\centering
		\begin{subfigure}[b]{0.4\textwidth}
			\centering
			\includegraphics[height=5.3cm,width=5cm]{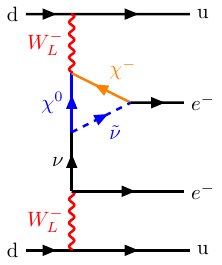}
			\caption{}
			\label{fig:3a}
		\end{subfigure}
		\hspace{35pt}
		\begin{subfigure}[b]{0.4\textwidth}
			\centering
			\includegraphics[height=5.3cm,width=5cm]{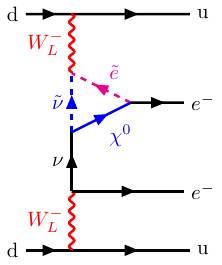}
			\caption{}
			\label{fig:3b}
		\end{subfigure}
		\caption{The SUSYLRM diagrams of the effective diagram 
			Fig.(\ref{fig:d}). The triangle loops consist 
			the SUSYLRM fields.}
		\label{fig:3rd_category}
\end{figure}

Following the same procedure as discussed in the section (\ref{sssec:2nd_category}). The particle physics parameters for the third category can be derived. The general effective vertex for the third category
is shown in Fig.(\ref{fig:d}) and the possible SUSYLRM diagrams
for this vertex are shown in Fig.(\ref{fig:3rd_category}).

Here, the $\eta$ terms for Fig.(\ref{fig:3a}) and 
Fig.(\ref{fig:3b}) are defined as,
	\begin{equation}
		\begin{split}
			& \eta_{3a} = (V_{ud})^2 V_{(e^- - W_L^- - \nu)} 
			V_{(\tilde{\nu} - \chi^0 - \nu)}. V_{(e^- - \chi^- - 
				\tilde{\nu})} V_{(W_L^- - \chi^0 - \chi^-)} \sum_{a=\tilde{\nu},
				\chi^0,\chi^-} 
			\frac{x_a^2 \ln{x_a}}{\prod_{a\neq b} (x_b - x_a)},
			\\& \eta_{3b} = (V_{ud})^2 V_{(e^- - W_L^- - \nu)}
			V_{(\tilde{\nu} - \chi^0 - \nu)} V_{(e^- - \chi^0 - \tilde{e})}
			V_{(W_L^- - \tilde{\nu} - \tilde{e})} \sum_{a=\tilde{e},
				\tilde{\nu},\chi^0} \frac{x_a^2 \ln{x_a}}{\prod_{a\neq b} 
				(x_b - x_a)}.
		\end{split}
	\end{equation}
So, from the third category, the total contribution is as
$\eta_3 = \eta_{3a} + \eta_{3b}$.

\subsubsection*{\bf \# Total Contribution from long-range interactions}	
\label{sec:Lrange}
Once the terms $\eta_2$ and $\eta_3$ are calculated from the
possible effective vertices, the amplitude for long-range 
interactions can be written as
	\begin{equation}
		\Lambda_{0\nu\beta\beta}^{lr} = \left<(A,Z+2),2e^- \left\lvert 
		\mathcal{T} \exp \left[i\int d^4 x \left(\frac{\eta_2}
		{\text{M}^2_R} J^{\mu} \bar{e} (\mathcal{O}_1)_{\mu}\nu_L^c 
		+ \eta_3 W^-_{\mu} \bar{e} (\mathcal{O}_2)^{\mu}
		\nu_L^c\right)\right] \right\lvert (A,Z) \right>.
		\label{eqn:amp_lr}
	\end{equation}
From this amplitude of long-range interactions, one can directly
write down the differential decay width following the Eq.(\ref{eqn:decay_width}). From this decay width formula, the half-life of the decay process
can be extracted as \cite{Ding:2024obt}, 
\begin{equation}
		\begin{split}
			\left(T_{1/2}^{0\nu\beta\beta}\right)^{-1} & = \frac{1}{\ln2}
			\int d\Gamma_{0\nu\beta\beta}^{(0_i^+ \rightarrow 0_f^+)} \\& = 
			\frac{1}{8\ln 2} \frac{1}{(2\pi)^5}\int \frac{d^3k_1}{2E_1} 
			\frac{d^3k_2}{2E_2} \lvert \Lambda_{0\nu\beta\beta}^{eff} 
			\lvert^2 F(Z,E_1)F(Z,E_2) \delta(E_1 + E_2 + E_f - E_i). 
\end{split}
\label{eqn:half-life_gen}
\end{equation}
Here, $F(Z, E)$ is the Fermi function, whose analytic form is given by,
\begin{equation}
F(Z,E) = \frac{4}{\lvert \Gamma(2\gamma_1 + 1) \lvert^2}
(2qR)^{2(\gamma_1 - 1)} \lvert \Gamma(\gamma_1 + iy) \lvert^2 
	\exp(\pi y),
\end{equation}
where $R$ is the nuclear radius, $R=r_0 A^{\frac{1}{3}}$ and
$\gamma_1 = \sqrt{1 - (\alpha Z)^2}$ and $y= \frac{\alpha Z E}
{q}.$ In the above expressions, $Z(A)$ is the atomic(nucleon)
number and $q_i$ is the electron momenta. Following the
expression of half-life in Eq.(\ref{eqn:half-life_gen}), 
it can be written in terms of the nuclear matrix element, phase space factor, and the dimensionless particle physics parameter,
which looks like
\begin{equation}
	\left(T_{1/2}^{0\nu\beta\beta}\right)^{-1}_{lr} = G_{01} \lvert 
		\mathcal{M}^{lr}_{\text{SUSYLRM}} \lvert^2  \lvert 
		\mathcal{\eta}_{\text{SUSYLRM}}^{lr} \lvert^2.
\label{eqn:halflife_lr}
\end{equation} 
Here, the phase space factor $G_{01}$ is defined as,
	\begin{equation}
		G_{01} = \frac{(G_F g_A)^4 m_e^2}{64 \pi^5 R^2 \ln2} \int
		F(Z,E_1)F(Z,E_2) \delta(E_1 + E_2 + E_f - E_i) q_1q_2E_1E_2
		dE_1 dE_2 d(\hat{q_1}.\hat{q_2}).
	\end{equation}

Here, the dimensionless particle physics parameter
$\mathcal{\eta}_{\text{SUSYLRM}}^{lr}$ is defined as, 
\begin{equation}
		\eta^{lr}_{\text{SUSYLRM}} = \eta_2 \left(\frac{m_p}
		{\text{M}_R}\right)^2 + \eta_3 \left(\frac{m_p}{m_W}\right)^2.
	\end{equation}
And $\mathcal{M}^{lr}_{\text{SUSYLRM}}$, the nuclear matrix element that
originates from the quark current, which is further updated into the nucleon
current by NRIA. This term usually defines the effect of initial and
final nuclear states and the finite structure of the nucleus. And the leptonic
part present in the amplitude influences the dimensionless particle physics
parameter $\eta$. The nuclear matrix element for this case is taken as,
	\begin{equation}
		\mathcal{M}_{\text{SUSYLRM}}^{lr} = \mathcal{M}_{S+P}^{S+P} + 
		\mathcal{M}_{T_L}^{T_R} + \mathcal{M}_{T_R}^{T_R}.
	\end{equation}   
	Here, each of the sub-NMEs is represented as,
	\begin{equation}
		\begin{split}
			&\mathcal{M}_{S+P}^{S+P} = \frac{F_p^3 (0)}{R_0 m_e f_A} 
			(\mathcal{M}_{T'} + \frac{1}{3}\mathcal{M}_{GT'}), \\&
			\mathcal{M}_{T_L}^{T_R} = -\alpha_1 \frac{2}{3}\mathcal{M}_{GT'} 
			+ \alpha_1 \mathcal{M}_T', \\&
			\mathcal{M}_{T_R}^{T_R} = \alpha_2 \mathcal{M}_F' 
			- \alpha_3 (\mathcal{M}_{T''} + \frac{1}{3}\mathcal{M}_{GT''}).	
		\end{split}
	\end{equation}
Here, $R_0$ is the nuclear radius, $m_e$ is the mass of electron and $f_A$
is the axial coupling constant. The coefficients $\alpha_1$, $\alpha_2$
and $\alpha_3$ are determined by the relations between the vector and axial
couplings which are shown in \cite{Pas:1999fc}. The values for these sub-NMEs 
are also taken from Ref. \cite{Pas:1999fc} and are shown in the Table.(\ref{tab:nmes}),
\begin{table}[htb]
\centering
	\setlength{\tabcolsep}{6pt} 
	\renewcommand{\arraystretch}{1}
	\begin{center}
			\begin{tabular}{|c |c| c| c| c|} 
				\hline
				$\mathcal{M}_{GT'}$ & $\mathcal{M}_{F'}$ & $\mathcal{M}_{GT''}$
				& $\mathcal{M}_{T'}$ & $\mathcal{M}_{T''}$  \\ [1.0ex] 
				\hline
				2.95 & -0.663 & 8.78 & 0.224 & 1.33  \\ 
				\hline
			\end{tabular}
			\caption{The sub-NME values are calculated through pn-QRPA in
				\cite{Muto:1989cd, Staudt:1990qi}.}
			\label{tab:nmes}
		\end{center}
\end{table}

\subsection{Fourth category: Short-range interactions}
\label{sssec:4th category} 
\begin{figure}[h]
		\centering
		\begin{subfigure}[b]{0.3\textwidth}
			\centering
			\includegraphics[width=\textwidth]{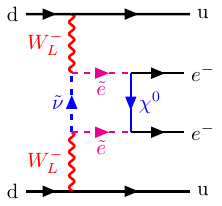}
			\caption{}
			\label{fig:4a}
		\end{subfigure}
		\hspace{10pt}
		\begin{subfigure}[b]{0.3\textwidth}
			\centering
			\includegraphics[width=\textwidth]{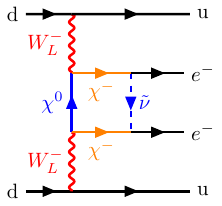}
			\caption{}
			\label{fig:4b}
		\end{subfigure}
		\hspace{10pt}
		\begin{subfigure}[b]{0.3\textwidth}
			\centering
			\includegraphics[width=.89\textwidth]{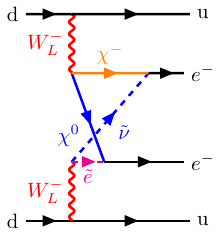}
			\caption{}
			\label{fig:4c}
		\end{subfigure}
		\caption{The SUSYLRM counterparts of the effective vertex 
			shown in Fig.(\ref{fig:e}).}
		\label{fig:4th_category}
	\end{figure}

Now exploring the short-range interactions, the first of 
them is shown in the effective vertex diagram Fig.(\ref{fig:e}).
The possible SUSYLRM extensions for this particular effective
vertex are shown in Fig.(\ref{fig:4th_category}).
	
And for Fig.(\ref{fig:4a}), (\ref{fig:4b}) and Fig.(\ref{fig:4c}) the corresponding particle physics parameters are,
\begin{equation}
		\begin{split}
			& \eta_{4a} = (V_{ud})^2 \left\{V_{(e^- - \chi^0 - \tilde{e})}
			V_{(W_L^- - \tilde{\nu} - \tilde{e})}\right\}^2 \sum_{a=\tilde{e},
				\tilde{\nu},\chi^0} \frac{x_a^2 \ln{x_a}}{\prod_{a\neq b} (x_b - x_a)},
			\\& \eta_{4b} = (V_{ud})^2 \left\{V_{(e^- - \chi^- - 
				\tilde{\nu})} V_{(W_L^- - \chi^0 - \chi^-)}\right\}^2 
			\sum_{a=\tilde{\nu},\chi^0,\chi^-} \frac{x_a^2 \ln{x_a}}
			{\prod_{a\neq b} (x_b - x_a)},
			\\& \eta_{4c} = (V_{ud})^2 V_{(W_L^- - \chi^0 - \chi^-)}
			V_{(e^- - \chi^- - \tilde{\nu})} V_{(e^- - \chi^0 - \tilde{e})}
			V_{(W_L^- - \tilde{\nu} - \tilde{e})} \sum_{a=\tilde{e},\tilde{\nu},
				\chi^0,\chi^-} \frac{x_a^2 \ln{x_a}}{\prod_{a\neq b} (x_b - x_a)}.
		\end{split}
	\end{equation}
	
The overall contribution from fourth category is, $\eta_4 = 
\eta_{4a} + \eta_{4b} + \eta_{4c}$.
\subsection{Fifth category: Short-range interactions}
\label{sssec:5th_category}
\begin{figure}[h]
\centering
\begin{subfigure}[b]{0.35\textwidth}\centering
\includegraphics[width=\textwidth]{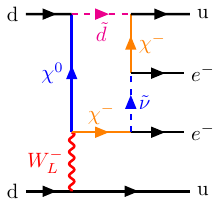}
\caption{}
\label{fig:5a}
\end{subfigure}
\hspace{10pt}
\begin{subfigure}[b]{0.35\textwidth}\centering
\includegraphics[width=.95\textwidth]{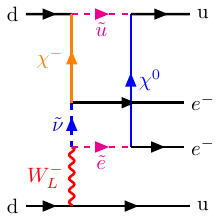}
\caption{}
\label{fig:5b}
\end{subfigure}
\hspace{10pt}
\begin{subfigure}[b]{0.35\textwidth}\centering
\includegraphics[width=.87\textwidth]{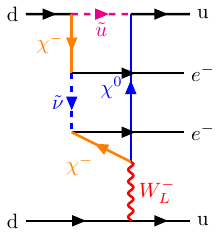}
\caption{}
\label{fig:5c}
\end{subfigure}
\hspace{10pt}
\begin{subfigure}[b]{0.35\textwidth}\centering
\includegraphics[width=\textwidth]{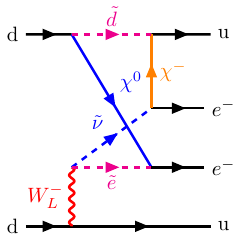}
\caption{}
\label{fig:5d}
\end{subfigure}
\caption{Possible SUSYLRM diagrams for Fig.(\ref{fig:f}).}
\label{fig:5th_category}
\end{figure}

The fifth category includes the maximum number of diagrams, 
which are shown in Fig.(\ref{fig:5th_category}). The effective 
vertex for this category is shown in Fig.(\ref{fig:f}) .
	
The particle physics parameters for the respective diagrams 
Fig.(\ref{fig:5a}), (\ref{fig:5b}), (\ref{fig:5c}) and 
(\ref{fig:5d}) are,
\begin{equation}
\begin{split}
& \eta_{5a} = V_{ud} (V_{(e^- - \chi^- - \tilde{\nu})})^2 
V_{(u - \tilde{d} - \chi^-)} V_{(d - \tilde{d} - \chi^0)} 
V_{(W_L^- - \chi^0 - \chi^-)} \sum_{a=\tilde{d}, \tilde{\nu}, 
\chi^0, \chi^-} \frac{x_a^2 \ln{x_a}}{\prod_{a\neq b} (x_b - x_a)},  
\\& \eta_{5b} =  V_{ud} V_{(d - \tilde{u} - \chi^-)} V_{(u -
\tilde{u} - \chi^0)} V_{(e^- - \chi^0 - \tilde{e})} V_{(W_L^- -
\tilde{\nu} - \tilde{e})} V_{(e^- - \chi^- - \tilde{\nu})} 
\sum_{a=\tilde{u}, \tilde{\nu}, \tilde{e}, \chi^0, \chi^-} 
\frac{x_a^2 \ln{x_a}}{\prod_{a\neq b} (x_b - x_a)},
\\& \eta_{5c} =  V_{ud} V_{(d - \tilde{u} - 
\chi^-)} V_{(u - \tilde{u} - \chi^0)} 
(V_{(e^- - \chi^- - \tilde{\nu})})^2 V_{(W_L^- - \chi^0 - 
\chi^-)} \sum_{a=\tilde{u}, \tilde{\nu}, \chi^0, \chi^-} 
\frac{x_a^2 \ln{x_a}}{\prod_{a\neq b} (x_b - x_a)},
\\& \eta_{5d} = V_{ud} V_{(e^- - W_L^- - \nu)}
V_{(\tilde{\nu} - \chi^0 - \nu)} V_{(e^- - \chi^- -
\tilde{\nu})} V_{(e^- - \chi^0 - \tilde{e})} V_{(W_L^- -
\tilde{\nu} - \tilde{e})} \sum_{a=\tilde{e}, \tilde{\nu},
\tilde{d}, \chi^-, \chi^0} \frac{x_a^2 \ln{x_a}}
{\prod_{a\neq b} (x_b - x_a)}.
\end{split}
\end{equation}
From the fifth class of diagrams, the total contribution comes as,
$\eta_5 = \eta_{5a} + \eta_{5b} + \eta_{5c} + \eta_{5d}$. 

\subsection{Sixth category: Short-range interactions}
\label{sssec:6th_category}
\begin{figure}[h]
\centering
\begin{subfigure}[b]{0.3\textwidth}
\centering
\includegraphics[height=5cm,width=5cm]{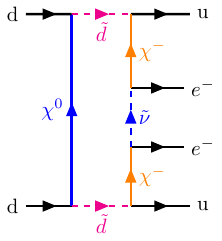}
\caption{}
\label{fig:6a}
\end{subfigure}
\hspace{10pt}
\begin{subfigure}[b]{0.3\textwidth}
\centering
\includegraphics[height=5cm,width=5cm]{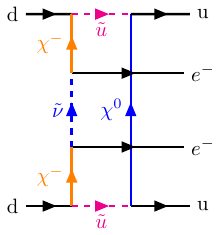}
\caption{}
\label{fig:6b}
\end{subfigure}
\hspace{10pt}
\begin{subfigure}[b]{0.3\textwidth}
\centering
\includegraphics[height=5cm,width=5cm]{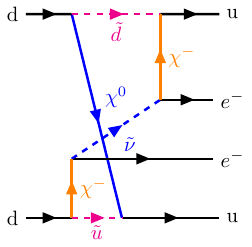}
\caption{}
\label{fig:6c}
\end{subfigure}
\caption{The diagrams coming from the effective vertex Fig.(\ref{fig:g}).}
\label{fig:6th_category}
\end{figure}

The last category includes three diagrams in Fig.(\ref{fig:6th_category}).
	
For Fig.(\ref{fig:6a}), (\ref{fig:6b}) and Fig.(\ref{fig:6c}) the $\eta$ terms are defined as,
\begin{equation}
\begin{split}
& \eta_{6a} = \left\{V_{(e^- - W_L^- - \nu)} V_{(\tilde{\nu} 
- \chi^0 - \nu)} V_{(e^- - \chi^- - \tilde{\nu})}\right\}^2
\sum_{a=\tilde{d}, \tilde{\nu}, \chi^-, \chi^0} \frac{x_a^2 
\ln{x_a}}{\prod_{a\neq b} (x_b - x_a)},
\\& \eta_{6b} = \left\{V_{(e^- - W_L^- - \nu)} 
V_{(d - \tilde{u} - \chi^-)} V_{(u - \tilde{u} - \chi^0)}
\right\}^2 \sum_{a=\tilde{u}, \tilde{\nu}, \chi^-, \chi^0}
\frac{x_a^2 \ln{x_a}}{\prod_{a\neq b} (x_b - x_a)},
\\& \eta_{6c} = \left(V_{(e^- - W_L^- - \nu)}\right)^2 
V_{(\tilde{\nu} - \chi^0 - \nu)} V_{(e^- - \chi^- - \tilde{\nu})}
V_{(d - \tilde{u} - \chi^-)} V_{(u - \tilde{u} - \chi^0)}
\sum_{a=\tilde{d}, \tilde{\nu}, \tilde{u}, \chi^-, \chi^0} 
\frac{x_a^2 \ln{x_a}}{\prod_{a\neq b} (x_b - x_a)}.
\end{split}
\end{equation}
	
The total contribution comes as, 
$\eta_6 = \eta_{6a} + \eta_{6b} + \eta_{6c}$.

\subsubsection*{\bf \# Total Contribution from short-range interactions}	
\label{sec:Srange}
Similarly, like the long-range case, here also the amplitude can 
be written as,
\begin{equation}
\begin{split}
\Lambda_{0\nu\beta\beta}^{\text{sr}} = \left< (A,Z+2),\, 2e^- \, \middle| \, 
\mathcal{T} \exp \left[ i \int d^4 x \left( 
\frac{\eta_4}{\text{M}_R} W_{\mu}^- W^-_{\nu}\bar{e}\, (\mathcal{O}_3)^{\mu \nu}
e^c \right. \right. \right. \\
\left. \left. \left.
+ \frac{\eta_5}{\text{M}^3_R} J^\mu W^-_\mu\, \bar{e}\,(\mathcal{O}_4)\, e^c 
+ \frac{\eta_6}{\text{M}^5_R} J^\mu J^\nu\, \bar{e}\,
(\mathcal{O}_5)_{\mu \nu}\, e^c
\right) \right] \, \middle| \, (A,Z) \right>.
\end{split}
\end{equation}
	
To evaluate the nuclear matrix elements (NMEs), 
it is necessary to convert the quark-level currents in the
Lagrangian into nucleon-level currents. This is typically 
done using the non-relativistic impulse approximation (NRIA).
Under this approximation, the nucleon current is expressed as:
\begin{equation}
J^{\mu}_{\text{nucleon}} = \sum_{i}\tau_+^i \left[(f_V - 
f_A C_i)g^{\mu0} - (f_A \sigma_i^k + f_V \vec{D}_i^k)g^{\mu k}
\right] \times \frac{m_A^3}{8\pi} e^{-m_A \lvert \vec{x} -
\vec{r}_i \rvert}
\end{equation}
Here, $f_V = 1$, $f_A = 1.261$, $m_A = 0.85$ GeV and $g^{\mu\nu}$ 
is the metric tensor. $\tau_+^i$ is the isospin-raising operator, 
and $\vec{r}_i$ is the position of the $i^{\text{th}}$ nucleon. 
The scalar and vector recoil terms are defined by $C_i$ and
$\vec{D}_i$, respectively, as:
\begin{equation}
\begin{split}
C_i &= \frac{1}{2m_p} \left[ (\vec{p}_i + \vec{p}'_i) \cdot 
\vec{\sigma}_i - \frac{f_P}{f_A}(E_i - E'_i)\vec{q}_i \cdot 
\vec{\sigma}_i \right], \\
\vec{D}_i &= \frac{1}{2m_p} \left[ (\vec{p}_i + 
\vec{p}'_i) + i\left(1 - 2m_p\frac{f_W}{f_V} \right)
\vec{q}_i \times \vec{\sigma}_i \right].
\end{split}
\end{equation}
In these expressions, $(\vec{p}_i, E_i)$ and $(\vec{p}'_i, E'_i)$
denote the initial and final three-momentum and energy of the 
$i^{\text{th}}$ nucleon, and the momentum transfer is defined 
as $\vec{q}_i = \vec{p}_i - \vec{p}'_i$. The relations between the couplings are given by:
\begin{equation}
\frac{f_W}{f_V} = -\frac{(\mu_p - \mu_n)}{2m_p}, \quad
\frac{f_P}{f_A} = \frac{2m_p}{m_{\pi}^2},
\end{equation}
where $m_\pi$ is the pion mass, and $\mu_p$, $\mu_n$ are the 
magnetic moments of the proton and neutron, respectively.
Now, incorporating the SUSYLRM effective vertices, the 
short-range amplitude for neutrinoless double beta decay can 
be written as,
\begin{equation}
\Lambda_{0\nu\beta\beta}^{\text{sr}} = \sqrt{2} C_{0\nu}^{-1} 
\left( \frac{\eta^{\text{sr}}_{\text{SUSYLRM}}}{\text{M}_R^5}
\right)
[\bar{e}(1 + \gamma_5)e^c] \left< 0_f^+ \lvert \Omega^{\text{sr}}_
{\text{SUSYLRM}} \lvert 0_i^+ \right>,
\end{equation}
where $C_{0\nu}$ is the normalization factor,
\begin{equation}
C_{0\nu} = \frac{4\pi}{m_p m_e} \frac{R_0}{f_A^2},
\end{equation}
with $R_0$ denoting the nuclear radius. The particle physics parameter in this case is:
\begin{equation}
\eta^{\text{sr}}_{\text{SUSYLRM}} = \eta_4 \left( 
\frac{\text{M}_R}{m_W} \right)^4 
+ \eta_5 \left( \frac{\text{M}_R}{m_W} 
\right)^2 + \eta_6.
\end{equation}
The nuclear transition dynamics are encoded in the operator
$\Omega^{\text{sr}}_{\text{SUSYLRM}}$, which gives rise to the
nuclear matrix element (NME) for short-range interactions:
\begin{equation}
\mathcal{M}_{\text{SUSYLRM}}^{\text{sr}} = \left< 0_f^+\lvert 
\Omega^{\text{sr}}_{\text{SUSYLRM}} \lvert 0_i^+ \right>.
\end{equation}
Based on the chosen benchmark points, all sparticle
masses are assumed to be heavy ($>100$~GeV). Hence, the nuclear matrix element can be approximated by the 
corresponding to heavy neutrino exchange,
\begin{equation}
\mathcal{M}_{\text{SUSYLRM}}^{\text{sr}} = \mathcal{M}_{F,N}
- \mathcal{M}_{GT,N},
\end{equation}
where the Fermi (F) and Gamow-Teller (GT) components are given 
by:
\begin{equation}
\mathcal{M}_{GT,N} = \left( \frac{m_A^2}{m_e m_p} \right) 
\left< 0_f^+ \lvert \Omega_{GT,N} \lvert 0_i^+ \right>, \quad
\mathcal{M}_{F,N} = \left( \frac{m_A^2}{m_e m_p} \right) 
\left( \frac{f_V}{f_A} \right)^2 \left< 0_f^+ \lvert \Omega_{F,N}
\lvert 0_i^+ \right>,
\end{equation}
with the transition operators defined as:
\begin{equation}
\begin{split}
\left< 0_f^+ \lvert \Omega_{GT,N} \lvert 0_i^+ \right> &= 
\left< 0_f^+ \left\lvert \sum_{i \ne j} \tau_+^i \tau_+^j \, 
\vec{\sigma}_i \cdot \vec{\sigma}_j \left( \frac{R_0}{r_{ij}} 
\right) F_N(x_A) \right\rvert 0_i^+ \right>, \\
\left< 0_f^+ \lvert \Omega_{F,N} \lvert 0_i^+ \right> &=
\left< 0_f^+ \left\lvert \sum_{i \ne j} \tau_+^i \tau_+^j 
\left( \frac{R_0}{r_{ij}} \right) F_N(x_A) \right\rvert 0_i^+ \right>.
\end{split}
\end{equation}
Here, $F_N(x_A)$ is the short-range potential given by,
\begin{equation}
F_N(x_A) = 4\pi m_A^6 r_{ij} \int \frac{d^3 \vec{q}}
{(2\pi)^3} \frac{e^{i \vec{q} \cdot \vec{r}_{ij}}}{(m_A^2 
+ \vec{q}^2)^4},
\end{equation}
where $x_A = m_A r_{ij}$. The analytic form of this potential
is,
\begin{equation}
F_N(x) = \frac{x}{48} (3 + 3x + x^2) e^{-x}.
\end{equation}
Using the proton-neutron Quasiparticle Random Phase Approximation
(pn-QRPA) model \cite{Pantis:1996py}, the nuclear matrix element is
computed for $\text{Ge}^{76}$ to be \cite{Hirsch:1996qw}:
\begin{equation}
\mathcal{M}^{\text{sr}}_{\text{SUSYLRM}} = 289.
\end{equation}
	
The corresponding inverse half-life for the short-range contribution can be calculated as \cite{Pas:2000vn},
\begin{equation}
\left( T_{1/2}^{0\nu\beta\beta} \right)^{-1}_{\text{sr}} = 
G_{01} \frac{4 m_p^2}{G_F^4} \left\lvert 
\frac{\eta^{\text{sr}}_{\text{SUSYLRM}}}{\text{M}_R^5} 
\mathcal{M}^{\text{sr}}_{\text{SUSYLRM}} 
\right\rvert^2.
\label{eqn:halflife_sr}
\end{equation}
\section{Numerical estimates and results}
\label{sec:result}
Using equations (\ref{eqn:halflife_lr}), and (\ref{eqn:halflife_sr})
the total inverse half-life for $0\nu\beta\beta$ decay in the 
SUSYLRM framework is obtained as the sum of short-range and 
long-range contributions. The total half-life expression is,
thus formulated as:
\begin{equation}
\begin{split}
\left(T_{1/2}^{0\nu\beta\beta}\right)_{\text{Total}} &= \left[
\left(T_{1/2}^{0\nu\beta\beta}\right)^{-1}_{\text{sr}} + 
\left(T_{1/2}^{0\nu\beta\beta}\right)^{-1}_{\text{lr}} 
\right]^{-1} \\
&= \left[ G_{01} \frac{4m_p^2}{G_F^4} \left| \frac{\eta^{
\text{sr}}_{\text{SUSYLRM}}}{\text{M}_R^5} 
\mathcal{M}^{\text{sr}}_{\text{SUSYLRM}} \right|^2 
+ G_{01} \left| \mathcal{M}^{\text{lr}}_{\text{SUSYLRM}}
\right|^2 \left| \eta^{\text{lr}}_{\text{SUSYLRM}} \right|^2
\right]^{-1}.
\end{split}
\end{equation}

The half-life expression depends on several parameters: the 
masses of neutralinos ($m_{\chi^0_i}$, $i = 1$ to $8$), 
charginos ($m_{\chi^\pm_i}$, $i = 1$ to $6$), squarks, 
charged sleptons, sneutrinos ($m_{\tilde{\nu}_i}$, $i = 1$ 
to $6$), and physical constants such as the electron mass 
$m_e$, proton mass $m_p$, the Fermi constant $G_F$, the 
phase space factor $G_{01}$, nuclear matrix elements, and
the SUSYLRM breaking scale $\text{M}_R$. The numerical value for phase factors results in units of ${year}^{-1}$, and for two widely used materials, Ge and Xe, the values are shown in Table (\ref{tab:phase_factor}). In our work, the value for Ge-76 is considered.
\begin{table}[htb]
\centering
\setlength{\tabcolsep}{6pt}
\renewcommand{\arraystretch}{1}
\begin{tabular}{|c|c|}
\hline
Isotope & $G_{01} ({year}^{-1})$  \\
\hline
Ge-76 & $5.77 \times 10^{-15}$  \\
\hline
Xe-136 & $3.56 \times 10^{-14}$  \\
\hline
\end{tabular}
\caption{The phase space factor values are shown in the above table \cite{Kotila:2012zza}.}
\label{tab:phase_factor}
\end{table}

The large dimensionality of this
parameter space poses a significant
challenge in obtaining phenomenologically consistent results.
This issue is efficiently resolved by employing well-motivated 
benchmark points (BPs) from the literature \cite{Frank:2017tsm,
Alloul:2013fra, PhysRevD.90.115021, Chatterjee:2018gca}. These
BPs have been constructed to satisfy existing collider bounds, 
flavor constraints, and dark-matter relic density bounds, while
maintaining consistency with theoretical requirements such as
vacuum stability and correct symmetry-breaking patterns.
In this work, we utilize these BPs to scan the relevant parameter
space effectively, enabling us to calculate the half-life of 
$0\nu\beta\beta$ decay in a robust and experimentally viable
setting.

Each benchmark point offers a unique balance between the
parameters while adhering to experimental constraints. A few
parameters remain fixed across all BPs, including:
\[\lambda_L = 0.4, \quad \lambda_S = -0.5,
\quad T_R = T_S = -2~\text{TeV}, \quad T_3 = 1~\text{TeV},\]
\[M_3 = 3.5~\text{TeV}, \quad M_{\Delta_{1L}}^2 = M_{\Delta_{2L}}^2
= 2~\text{TeV}^2, \quad \zeta_F = -5 \times 10^5~\text{GeV}^2,
\]\[(Y_L^4)_{ii} = (0.019, 0.022, 0.1).\]
	
\begin{figure}[htb]
\centering
\includegraphics[width=16.4cm, height=11.5cm]{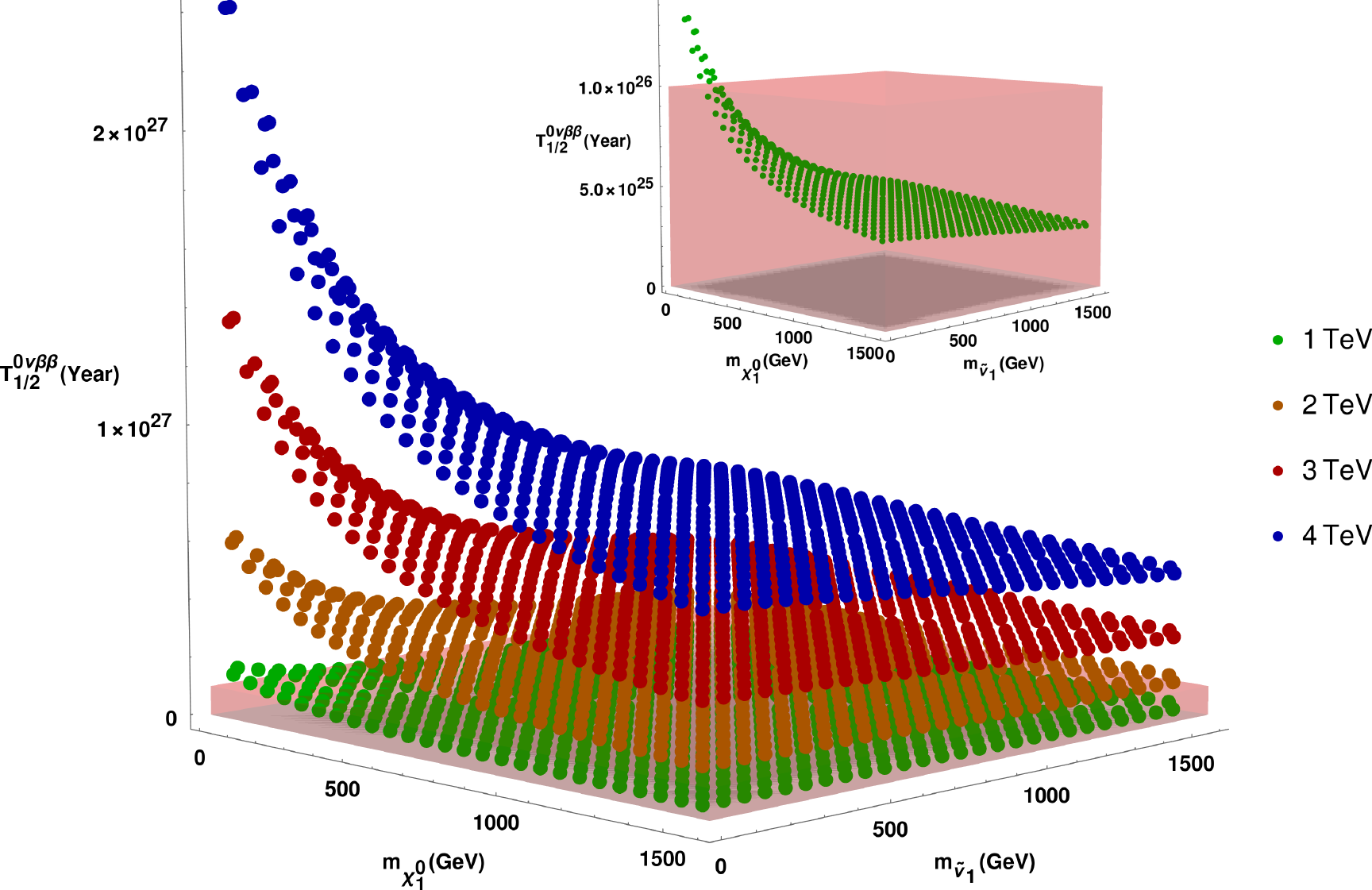}
\caption{The half-life $T_{1/2}^{0\nu\beta\beta}$ for 
neutrinoless double beta decay is plotted against the 
masses of the lightest neutralino ($m_{\chi^0_1}$) and 
sneutrino ($m_{\tilde{\nu}_1}$). Different colored dots 
surfaces correspond to different SUSYLRM breaking scales
$\text{M}_R = 1, 2, 3$, and $4$ TeV, shown in green, orange, red, and blue, respectively. The red box is the volume excluded by the current experimental results.}
\label{fig:t_half_SUSYLR}
\end{figure}

The values of the parameters specific to each benchmark point are
summarized in Table~\ref{tab:npv}. The results shown in Fig.(\ref{fig:t_half_SUSYLR}) are calculated for BP2. Although we report the results using BP2, the rest of the BPs result in similar outcomes. This is expected for almost closely lying values for the BPs. 
\begin{table}[htb]
\centering
\setlength{\tabcolsep}{6pt}
\renewcommand{\arraystretch}{1}
\begin{tabular}{|c |c| c| c| c| c|}
\hline
Parameters & BP1 & BP2 & BP3 & BP4 & BP5 \\
\hline
$\tan \beta$ & 8 & 7 & 7 & 7 & 7 \\
\hline
$\tan \beta_R$ & 1.05 & 1.045 & 1.045 & 1.04 & 1.045 \\
\hline
$v_S$ (TeV) & 10 & 7.2 & 6.4 &3 7.8 & 7 \\
\hline
$\lambda_3$ & 0.10 & 0.14 & 0.091 & 0.12 & 0.144 \\
\hline
$\lambda_R$ & 0.85 & 0.90 & 0.90 & 0.90 & 0.90 \\
\hline
$M_1$ (GeV) & 400 & 700 & 550 & 750 & 700 \\
\hline
$M_{2L}$ (GeV) & 900 & 1000 & 900 & 412 & 1200 \\
\hline
$M_{2R}$ (GeV) & 900 & 1000 & 900 & 1100 & 650 \\
\hline
\end{tabular}
\caption{Benchmark values used in the numerical 
calculation of the $0\nu\beta\beta$ decay half-life, 
consistent with Ref.\cite{Chatterjee:2018gca}.}
\label{tab:npv}
\end{table}
	
A more extensive list of sparticle masses and other relevant parameters for each BP can be 
found in Ref.\cite{Chatterjee:2018gca},
from which our numerical inputs are taken. An important phenomenological aspect
of the SUSYLRM is the role of the lightest neutralino and sneutrino as potential 
dark-matter candidates. Hence, any predictive result for 
$0\nu\beta\beta$ decay within this framework could also shed 
light on possible dark-matter contributions to low-energy processes. 
After fixing all the parameters according to the chosen BPs,
we plot the total half-life $T_{1/2}^{0\nu\beta\beta}$ as a 
function of the lightest sneutrino and neutralino masses, as
shown in Fig.(\ref{fig:t_half_SUSYLR}).
	
To comprehensively cover the parameter space, the lightest 
neutralino and sneutrino masses ($m_{\chi^0_1}$, 
$m_{\tilde{\nu}_1}$) are varied from $0$ to $1.5\times 10^3$ GeV. 
The parity breaking scale $\text{M}_R$ is set at
four reference values: $1, 2, 3$, and $4$ TeV. As illustrated 
in Fig.(\ref{fig:t_half_SUSYLR}), the predicted half-life strongly 
depends on these scales, providing a clear phenomenological 
handle on the role of supersymmetric contributions in 
$0\nu\beta\beta$ decay.
		
\subsection{Discussion}
\label{subsec:disc}
The results obtained so far can be analysed, keeping in mind the potential experimental searches for 
$0\nu\beta\beta$ decay and supersymmetry.
Several experimental collaborations worldwide are searching for the $0\nu\beta\beta$ decay, but no direct signal has been observed to date. However, each of the current setups provides a constraint on the half-life of
$0\nu\beta\beta$ decay, and they almost yield the same order of magnitude, for the lower limit ($\mathcal{O}(10^{26})$ Years). In the near future, experiments are expected to raise the lower bound on the half-life to 
$\mathcal{O}(10^{28})$ $years$. 
In the table (\ref{tab:onbb_expt}) the current and future experiments, along with their results, are shown for reference.
\begin{table}[htb!]
\centering
\setlength{\tabcolsep}{6pt}
\renewcommand{\arraystretch}{1}
\begin{tabular}{|c |c |c |}
\hline
Experiment & Lower limit of $T_{1/2}^{0\nu\beta\beta} (\text{Year})$  & Allowed scales of $\text{M}_R$ \\
\hline
KAMLAND-Zen \cite{KamLAND-Zen:2022tow} & $> 2.3\times 10^{26}$ & \orangedot{} \reddot{} \bluedot{} \\
\hline
GERDA \cite{GERDA:2020xhi} & $> 1.8\times 10^{26}$ & \orangedot{} \reddot{} \bluedot{}\\
\hline
EXO-200 \cite{EXO-200:2019rkq} & $> 3.5\times 10^{25}$ & \greendot{} \orangedot{} \reddot{} \bluedot{}\\
\hline
CUPID-Mo \cite{Armengaud:2019loe} & $> 2.43\times 10^{24}$ & \greendot{} \orangedot{} \reddot{} \bluedot{}\\
\hline
$\text{nEXO}^*$ \cite{nEXO:2021ujk} & $> 1.35\times 10^{28}$ & $>$ \bluedot{}\\
\hline
$\text{LEGEND}^*$ \cite{LEGEND:2017cdu} & $> 10^{28}$ & $>$ \bluedot{}\\
\hline
\end{tabular}
\caption{The above table shows the results from current and future experimental collaborations for detecting the $0\nu\beta\beta$ decay. The future experiments are marked with an asterisk (*). The allowed values of $\text{M}_R$ are shown in the third column using the color codes. The green, orange,
red and blue dots correspond to $M_R = 1, 2,3$ and $4$ TeV respectively.}
\label{tab:onbb_expt}
\end{table}

Although the discovery of the Higgs boson \cite{ATLAS:2012yve, CMS:2012qbp} had a significant impact on the validity of the SM, there is no experimental evidence of any SUSY particles yet. Nevertheless, there are substantial constraints on SUSY particles in terms of their masses that one can use in a phenomenological study. We want to highlight the results of GAMBIT collaboration \cite{GAMBIT:2018gjo}, which utilizes the ATLAS and CMS data with $36$ $\text{fb}^{-1}$ of p-p collision events. They point out a wide range of masses for neutralinos, $(8 - 500)$ GeV, and charginos, $(100 - 500)$ GeV at $95\%$ ${\rm C.L.}$.
With time, the search program has benefited from increased integrated luminosity and higher center-of-mass energy, which have helped to gain sensitivity to a wide range of the MSSM parameter space. The current data from ATLAS \cite{ATLAS:2024lda, ATLAS:2024woy} and CMS \cite{CMS-PAS-SUS-23-014} are based on events up to $140$ $\text{fb}^{-1}$ of p-p collisions at $\sqrt{s}$ = $13$ TeV. ATLAS (CMS) shows the minimum exclusion limit for the mass of neutralino LSP is around $100 (50)$ GeV, where, depending on the decay channel, the limit may increase up to $2000 (1500)$ GeV. For charginos, the observed exclusion limit from ATLAS ranges up to $(800 - 1000)$ GeV, while the CMS exclusion range is up to $(800 - 1200)$ GeV.


Now, coming to our analysis of the half-life of $0\nu\beta\beta$ in SUSYLRM,
by varying the parity-breaking scale in the range, $\text{M}_R \simeq 1-4$ TeV, 
we obtain distinct solution surfaces in the $T_{1/2}^{0\nu\beta\beta}$
vs. $(m_{\chi_1^0}, m_{\tilde{\nu}_1})$ parameter space $\text{M}_R \simeq 1,2 ,3$ and $4$ TeV, as shown in 
Fig.(\ref{fig:t_half_SUSYLR}). These are represented by green, orange,
red and blue dot surfaces, respectively. Except for a small 
region of the green surface (corresponding to the lowest breaking 
scale), the predicted half-lives lie above the present experimental 
lower bounds \cite{KamLAND-Zen:2022tow, Shtembari:2022pfs}, ensuring
phenomenological viability. An important
observation is that the half-life tends to be larger for lower 
values of $m_{\chi_1^0}$ and $m_{\tilde{\nu}_1}$, and decreases with
increase in their masses until it saturates at high mass values, a trend 
captured by the valley-like region in the 3D plot. Moreover, 
by increasing the parity-breaking scale, the value of the predicted half-life 
shifts to higher values, reinforcing its role as a crucial control parameter
in the theory.	These results suggest that, for fixed values of other
SUSYLRM parameters, smaller values of $m_{\chi_1^0}$ and 
$m_{\tilde{\nu}_1}$ (specifically, below $500$ GeV) can yield a half-life
predictions within the experimentally accessible range. Such mass 
values are consistent with the notion of lightest supersymmetric 
particles and support the interpretation of sneutrino and neutralino
as a viable dark matter candidate. Across all benchmarks 
scenarios, the qualitative behavior remains consistent, with only 
minor quantitative differences. However, with continuous updates of theoretical, experimental, and cosmological constraints, the BPs may change by a significant amount. 
This would lead to changes in the plot as shown in Fig.(\ref{fig:t_half_SUSYLR}) in terms 
allowed values of the parameters. A summary of the 
results representing the allowed values of $M_R$ corresponding to allowed
lower-limit on
half-life of $0\nu\beta\beta$ is presented in table (\ref{tab:onbb_expt})
where the green, orange,
red and blue dots correspond to $M_R = 1, 2,3$ and $4$ TeV respectively. For the largest considered breaking
scale of $\text{M}_R = 4$ TeV, the predicted half-life 
approaches $3 \times 10^{27}$ years. This value, while currently
allowed, may be challenged by next-generation $0\nu\beta\beta$ 
experiments targeting sensitivity up to $10^{28}$ years 
\cite{nEXO:2017nam, LEGEND:2017cdu, Armengaud:2019loe}. 
Any future non-observation in that sensitivity would potentially rule out lower
parity-breaking scales in SUSYLRM and push the preferred breaking scale into the multi-TeV
regime.
\section{Conclusion}
\label{sec:discussion}
The $0\nu\beta\beta$ decay is a promising probe for physics beyond the SM.
By measuring the helicities of the outgoing electrons, if it is found to be right-handed,
then it is a clear signal of new physics in support of left-right symmetry.
However, even if both electrons are found to be left-handed, there may be new 
physics present at the vertex level that can induce the decay.
In this study, we have pursued a systematic and minimal effective
field theory approach to calculate the half-life of neutrinoless
double beta decay within the framework of a supersymmetric left-right model. Starting from the effective Lagrangian, the contributions were classified into 
long- and short-range interactions. For each category, 
all relevant effective vertices were considered, and their 
contributions were computed and expressed in terms of dimensionless
particle physics parameters ($\eta$'s). These $\eta$ parameters 
were then combined with the nuclear matrix elements that 
encode the nuclear structure effects of the initial and final states.
Following the established methodology for $0\nu\beta\beta$ decay,
the total half-life expression was derived in terms of SUSYLRM
parameters. A notable feature of this formulation is the explicit
dependence on the parity-breaking scale $\text{M}_R$
which makes the half-life calculation sensitive to new physics 
effects from SUSYLRM. 
Furthermore, the role of dark matter
candidates in particular, the lightest sneutrino and neutralino
has been incorporated into the analysis, highlighting the interplay 
between $0\nu\beta\beta$ decay and supersymmetric DM phenomenology.

In conclusion, this analysis demonstrates that $0\nu\beta\beta$ decay 
provides a promising probe of the scale of parity breaking in a supersymmetric left-right model, 
with the potential to explore both the mechanism of neutrino mass
generation and the properties of a supersymmetric dark matter candidate. The
interplay between model parameters, especially the parity-breaking 
scale and the masses of light supersymmetric particles, offers a
promising phenomenological window that can be further constrained
by upcoming experimental results.
\section{Acknowledgements}
The authors thank Srubabati Goswami for proposing this study.
VB sincerely thanks the Ministry of Education, 
Government of India, for the financial assistance.
\section{Appendix}    
\subsection{Charginos and neutralinos of SUSYLRM}
\label{app:gauginos_SUSYLR}
In SUSY models, all gauginos have a particularly defined mass, which
introduces mixing between the gaugino states. In general, charged and
neutral gauginos are expressed separately as charginos ($\chi^{\pm}$)
and neutralinos ($\chi^0$). The SUSYLRM model includes six singly-charged
charginos, defined as $\chi_i^{\pm}$, where ($i$= 1 to 6). They are formed
in the basis $(\tilde{\Delta}_L^{\pm}, \tilde{\Delta}_R^{\pm}, 
\tilde{\Phi}_1^{\pm}, \tilde{\Phi}_2^{\pm}, \tilde{W}_L^{\pm}, 
\tilde{W}_R^{\pm})$,
where the mass matrix is expressed as
\begin{equation}
M_{\tilde{\chi}^{\pm}} = \begin{pmatrix}
\frac{\lambda_L v_s}{\sqrt{2}} & 0 & 0 & 0 & 0 & 0 \\
0 & \frac{\lambda_L v_s}{\sqrt{2}} & 0 & 0 & 0 & -g_Rv_{1R} \\
0 & 0 & 0 & \mu_{\text{eff}} & \frac{g_Lv_u}{\sqrt{2}} & 0 \\
0 & 0 & \mu_{\text{eff}} & 0 & 0 & -\frac{g_Rv_d}{\sqrt{2}} \\
0 & 0 & 0 & \frac{g_Lv_d}{\sqrt{2}} & M_{2L} & 0 \\
0 & g_Rv_{2R} & -\frac{g_rv_u}{\sqrt{2}} & 0 & 0 & M_{2R}
\end{pmatrix},
\label{eqn:chargino_mass-matrix}
\end{equation}
where $v_d = v\cos\beta$ and $v_u = v\sin\beta$.
When the left triplet and two neutral bidoublet Higgs bosons are inert, the neutralino sector consists of twelve fields. The twelve-dimensional mass matrices can be separated into three block-diagonal matrices. The left triplet and bidoublet mass matrices
are expressed as
\begin{equation}
M_{\tilde{\chi}\phi} = \begin{pmatrix}
0 & -\mu_{\text{eff}} \\
-\mu_{\text{eff}} & 0
\end{pmatrix}, \quad M_{\tilde{\chi}\delta} = \begin{pmatrix}
0 & \mu_L \\
\mu_L & 0
\end{pmatrix},
\end{equation} 
in the bases ($\tilde{\phi}_2, \tilde{\phi}_1$) and 
 ($\tilde{\delta}_{1L}, \tilde{\delta}_{2L}$) respectively.	
The neutralino mass matrix is given by
\begin{equation}
M_{\tilde{\chi}^0} = \begin{pmatrix}
0 & -\mu_{\text{eff}} & 0 & 0 & -\mu_d & 0 & \frac{g_Lv_u}
{\sqrt{2}} & -\frac{g_Rv_u}{\sqrt{2}} \\
-\mu_{\text{eff}} & 0 & 0 & 0 & -\mu_u & 0 & -\frac{g_Lv_d}
{\sqrt{2}} & \frac{g_Rv_d}{\sqrt{2}} \\
0 & 0 & 0 & \mu_R & \frac{\lambda_Rv_{2R}}{\sqrt{2}} & g'v_{1R}
& 0 & -g_Rv_{1R} \\
0 & 0 & \mu_R & 0 & \frac{\lambda_Rv_{1R}}{\sqrt{2}} & -g'v_{2R}
& 0 & -g_Rv_{2R} \\
-\mu_d & -\mu_u & \frac{\lambda_Rv_{2R}}{\sqrt{2}} & 
\frac{\lambda_Rv_{1R}}{\sqrt{2}} & \mu_S & 0 & 0 & 0 \\
0 & 0 & g'v_R & -g'v_{2R} & 0 & M_1 & 0 & 0 \\
\frac{g_Lv_u}{\sqrt{2}} & -\frac{g_Lv_d}{\sqrt{2}} & 0 & 0 & 
0 & 0 & M_{2L} & 0 \\
-\frac{g_Rv_u}{\sqrt{2}} & \frac{g_Rv_d}{\sqrt{2}} & -g_Rv_{1R} 
& -g_Rv_{2R} & 0 & 0 & 0 & M_{2R}
\end{pmatrix},
\label{eqn:neutralino_mass-matrix}
\end{equation}
in the basis, $(\tilde{\phi_1}, \tilde{\phi_2},\tilde{\delta}_{1R},
\tilde{\delta}_{2R},\tilde{S},\tilde{B},\tilde{W}_L^0,\tilde{W}_R^0)$. 
In the above matrices, the $\mu$ terms are expressed as $\mu_L = 
\lambda_L \frac{v_S}{\sqrt{2}}$, $\mu_R = \lambda_R \frac{v_S}{\sqrt{2}}$,
$\mu_u = \lambda_3 \frac{v_u}{\sqrt{2}}$, $\mu_d = \lambda_3 
\frac{v_d}{\sqrt{2}}$ and $\mu_S = \lambda_S \frac{v_S}{\sqrt{2}}$. 
For the gauge groups $SU(2)_L$, $SU(2)_R$ and $U(1)_{B-L}$ the
couplings are $g_L$, $g_R$ and $g_{B-L}$ respectively.
	
The chargino and neutralino mass matrices are diagonalized using the
matrices,
\begin{equation}
Z^{\dagger}_{\tilde{\chi}^-} M_{\tilde{\chi}^{\pm}} Z_{\tilde{\chi}^+}
= \text{Diag}(m_{\tilde{\chi}_1^{\pm}}, m_{\tilde{\chi}_2^{\pm}},
m_{\tilde{\chi}_3^{\pm}}, m_{\tilde{\chi}_4^{\pm}}, 
m_{\tilde{\chi}_5^{\pm}}, m_{\tilde{\chi}_6^{\pm}}), 
\end{equation}
where, $\tilde{\chi}^{\pm}_i = (Z_{\tilde{\chi}^{\pm}})_{i1} 
\tilde{\Delta}_L^{\pm} + (Z_{\tilde{\chi}^{\pm}})_{i2}
\tilde{\Delta}_R^{\pm} + (Z_{\tilde{\chi}^{\pm}})_{i3} 
\tilde{\Phi}_1^{\pm} + (Z_{\tilde{\chi}^{\pm}})_{i4} 
\tilde{\Phi}_2^{\pm} + (Z_{\tilde{\chi}^{\pm}})_{i5}W_L^{\pm}
+ (Z_{\tilde{\chi}^{\pm}})_{i6}W_R^{\pm}$.
And the neutralino mass matrix can be diagonalized by the matrix
$Z_{\tilde{\chi}^0}$ as
\begin{equation}
Z^{\dagger}_{\tilde{\chi}^0} M_{\tilde{\chi}^0} Z_{\tilde{\chi}^0} 
= \text{Diag}(m_{\tilde{\chi}_1^0}, m_{\tilde{\chi}_2^0},
m_{\tilde{\chi}_3^0}, m_{\tilde{\chi}_4^0}, m_{\tilde{\chi}_5^0},
m_{\tilde{\chi}_6^0}, m_{\tilde{\chi}_7^0}, m_{\tilde{\chi}_8^0}). 
\end{equation}
where, $\tilde{\chi}^0_i = (Z_{\tilde{\chi}^0})_{i1}\tilde{\phi_1}
+ (Z_{\tilde{\chi}^0})_{i2}\tilde{\phi_2} + (Z_{\tilde{\chi}^0})_{i3}
\tilde{\delta}_{1R} + (Z_{\tilde{\chi}^0})_{i4}\tilde{\delta}_{2R} + 
(Z_{\tilde{\chi}^0})_{i5}\tilde{S} + (Z_{\tilde{\chi}^0})_{i6}
\tilde{B} + (Z_{\tilde{\chi}^0})_{i7}\tilde{W}_L^0 + 
	Z_{\tilde{\chi}^0})_{i8}\tilde{W}_R^0$.
	
\subsection{Sleptons and squarks of SUSYLRM}
\label{app:sleptons}
Denoting the charged sleptons as $\tilde{L}_L^i = (\tilde{e},
\tilde{\mu}, \tilde{\tau})_L$ and $\tilde{L}_R^i = (\tilde{e}, 
\tilde{\mu}, \tilde{\tau})_R$, the mass-squared matrix for them
is expressed as, 
\begin{equation}
M_{\tilde{L}}^2 = \begin{pmatrix}
m_{\tilde{L}_L}^2 + m_l^2 + D_{11} & (T^3_L)_{ij}v\cos\beta + 
\mu_{\text{eff}}m_l \tan\beta \\
(T^3_L)_{ij}v\cos\beta + \mu_{\text{eff}}m_l \tan\beta & 
m_{\tilde{L}_R}^2 + m_l^2 + D_{22}
\end{pmatrix},
\end{equation}
where, $\mu_{\text{eff}} = \lambda_3v_s/\sqrt{2}$ and the forms
of D-terms are
\begin{equation}
\begin{split}
& D_{11} = -\frac{g_L^2}{8}v^2 \cos2\beta + g_{B-L}^2 (v_{1R}^2 
- v_{2R}^2), \\& 
D_{22} = \frac{g_R^2}{8}[2(v_{1R}^2 - v_{2R}^2) 
- v^2 \cos 2\beta] - g_{B-L}^2(v_{1R}^2 - v_{2R}^2).
\end{split}
\end{equation}
	
The sneutrino sector consists of two parts: scalar and pseudoscalar. 
The scalar mass matrix in $(\tilde{\nu}_L, \tilde{\nu}_R)$ basis is
\begin{equation}
M_{\tilde{\nu}}^2 = \begin{pmatrix}
m_{\tilde{L}_L}^2 + D_{11} & \left((T_L^2 v - Y_L^2 Y_L^4 v_{1R})
\sin\beta + Y_L^2 \mu_{\text{eff}}\frac{v\cos\beta}{\sqrt{2}}\right) 
\\ \left((T_L^2 v - Y_L^2 Y_L^4 v_{1R})\sin\beta + Y_L^2 
\mu_{\text{eff}}\frac{v\cos\beta}{\sqrt{2}}\right) & m_{\tilde{L}_R}^2 
+ D_{22} - \sqrt{2}T_L^4 v_{1R} + Y_L^4 (2 Y_L^4 v_{1R}^2 + 
\lambda_R v_S v_{2R})
\end{pmatrix},
\end{equation}
	
and the pseudo-scalar sneutrino mass matrix is
\begin{equation}
M_{\tilde{\nu}_{\text{p}}}^2 = \begin{pmatrix}
m_{\tilde{L}_L}^2 + D_{11} & \left((T_L^2 v + Y_L^2 Y_L^4 v_{1R})
\sin\beta + Y_L^2 \mu_{\text{eff}}\frac{v\cos\beta}{\sqrt{2}}\right)
\\ \left((T_L^2 v + Y_L^2 Y_L^4 v_{1R})\sin\beta + Y_L^2 
\mu_{\text{eff}}\frac{v\cos\beta}{\sqrt{2}}\right) & m_{\tilde{L}_R}^2
+ D_{22} + \sqrt{2}T_L^4 v_{1R} + Y_L^4 (2 Y_L^4 v_{1R}^2 -
\lambda_R v_S v_{2R})
\end{pmatrix}.
\end{equation}
	
Like sleptons, the squark sector has mass matrices for up and down type squarks in the bases $(\tilde{u}_L,
\tilde{u}_r)$ and $(\tilde{d}_L, \tilde{d}_R)$ respectively. 
The mass-squared matrices for them are
\begin{equation}
M_{\tilde{d}}^2 = \begin{pmatrix}
m_{\tilde{d}_L \tilde{d}_L^*} & m_{\tilde{d}_L \tilde{d}_R^*} \\
m^{\dagger}_{\tilde{d}_L \tilde{d}_R^*} & m_{\tilde{d}_R \tilde{d}_R^*}
\end{pmatrix}, \quad
M_{\tilde{u}}^2 = \begin{pmatrix}
m_{\tilde{u}_L \tilde{u}_L^*} & m_{\tilde{u}_L \tilde{u}_R^*} \\
m^{\dagger}_{\tilde{u}_L \tilde{u}_R^*} & m_{\tilde{u}_R \tilde{u}_R^*}
\end{pmatrix}
\label{eqn:squarks}
\end{equation}
	
Like the chargino and neutralino sector, the sfermion squared mass 
matrices are also diagonalized by the unitary matrices $Z_{\tilde{L}}, 
Z_{\tilde{\nu}}, Z_{\tilde{\nu_{\text{p}}}}, Z_{\tilde{d}}$ and 
$Z_{\tilde{u}}$ for the squared mass matrices $M_{\tilde{L}}^2,
M_{\tilde{\nu}}^2, M_{\tilde{\nu}_{\text{p}}}^2, M_{\tilde{d}}^2$
and $M_{\tilde{u}}^2$ respectively. To get the complete mass matrices
shown in Eq.(\ref{eqn:squarks}) and a detailed study of the Higgs sector
of SUSYLRM, we would refer to the article \cite{Bhattacherjee:2018xzf}.

\subsection{Sub-interactions used in this study}
\label{app:sub_interactions}
Ten simple interactions have contributed
to the complete study. All the loops present in the diagrams can be
decomposed into these simple interaction vertices. The vertex factors for
all the simple interactions are shown below. We can reformulate these
factors for different SUSY models, following the basics of SUSY vertices
\cite{Rosiek:1989sce, Drees:2004jm}. The vertex factors used in our study are in the vertices where SUSY fields and the mixing between SUSY gauge states are present. All of these mixings are introduced in the vertex factors with the help of mixing matrices, defined as $Z$'s. One can easily calculate the forms of the mixing matrices for sneutrinos ($Z_{\tilde{\nu}}$), squarks ($Z_{\tilde{d}}, Z_{\tilde{u}}$), neutralinos ($Z_{\tilde{\chi}^0}$), charginos ($Z_{\tilde{\chi}^{\pm}}$), etc. using the mass-squared matrices for sparticles as shown in the appendix (\ref{app:gauginos_SUSYLR} and \ref{app:sleptons}).
\begin{enumerate}
    \item $(e^- - W_L^- - \nu)$\\
    $\rightarrow$ $V_{\left(e^- - W_L^- - \nu\right)} = \frac{-ig}
{2\sqrt{2}}\gamma_{\mu}(1-\gamma_5)$.

\item {$(\tilde{\nu} - \chi^0 - \nu)$}\\
$\rightarrow$ $V_{\left(\tilde{\nu} - \chi^0 - \nu\right)} 
= \frac{ie}{\sqrt{2}s_Wc_W}Z_{\tilde{\nu}}^{IJ*}(Z_{\nu}^{1i}s_W - 
Z_{\nu}^{2j}c_W )P_L$.
	
\item {$(e^- - \chi^- - \tilde{\nu})$}\\
$\rightarrow$ $V_{\left(e^- - \chi^- - \tilde{\nu}\right)} = 
-i\left(\frac{e}{s_W}Z_{\tilde{\chi}^+}^{1i}P_L + 
Y_L^IZ_{\tilde{\chi}^-}^{21*}P_R\right)
Z_{\tilde{\nu}}^{IJ*}$.
	
\item {$(u - \tilde{d} - \chi^-)$}\\
$\rightarrow$ $V_{\left(u - \tilde{d} - \chi^-\right)} =
i\left[ -\left( \frac{e}{s_W}Z_{\tilde{d}}^{Ii}Z_{\tilde{\chi}^-}^{1j} + 
Y_Q^IZ_{\tilde{d}}^{(I+3)i}Z_{\tilde{\chi}^-}^{2j} \right)P_L  + Y_Q^J 
Z_{\tilde{d}}^{Ii}Z_{\tilde{\chi}^+}^{2j*}
P_R\right]K^{IJ*}$.
	
\item {$(d - \tilde{d} - \chi^0)$}\\
\begin{eqnarray*}
\hspace{-27mm}\rightarrow V_{\left(d - \tilde{d} - \chi^0\right)} &=& i\bigg[\left( \frac{-e}{\sqrt{2}s_Wc_W}Z_{\tilde{d}}^{Ii}(\frac{1}{3}Z_{\tilde{\nu}}^{1i}s_W - 
Z_{\tilde{\nu}}^{2j}c_W) + Y_Q^IZ_{\tilde{d}}^{i(I+3)}Z_{\tilde{\nu}}^{3j}\right)P_L 
\nonumber \\ &&+\left(\frac{-e\sqrt{2}}{3c_W}Z_{\tilde{d}}^{i(I+3)}Z_{\tilde{\nu}}^{1j*}
+ Y_Q^IZ_{\tilde{d}}^{Ii}Z_{\tilde{\nu}}^{3j*}\right)P_L \bigg].
\end{eqnarray*}
	
\item {$(d - \tilde{u} - \chi^-)$}\\
$\rightarrow V_{\left(d - \tilde{u} - \chi^-\right)} =$ $i\left[ \left(\frac{-e}{s_W}Z_{\tilde{u}}^{Ji*}Z_{\tilde{\chi}^+}^{1j} + Y_Q^J Z_{\tilde{u}}^{i(J+3)*}Z_{\tilde{\chi}^+}^{2j}\right)P_L - Y_Q^I Z_{\tilde{u}}^{Ji*}Z_{\tilde{\chi}^-}^{2j*}P_R\right]K^{JI}$.
	
\item {$(u - \tilde{u} - \chi^0)$}\\
\begin{eqnarray*}
\hspace{-25mm}	\rightarrow V_{\left(u - \tilde{u} - \chi^0\right)}  &=& i\bigg[\left(\frac{-e}{\sqrt{2}s_Wc_W}Z_{\tilde{u}}^{Ii*}(\frac{1}{3}Z_{\tilde{\nu}}^{1j}s_W + Z_{\tilde{\nu}}^{2j}c_W) - Y_Q^IZ_{\tilde{u}}^{i(I+3)*}Z_{\tilde{\nu}}^{4j}\right)P_L  \nonumber \\
&&+ \left(\frac{2\sqrt{2}}{3c_W}Z_{\tilde{u}}^{i(I+3)*}Z_{\tilde{\nu}}^{1*} - Y_Q^IZ_{\tilde{u}}^{Ii*}Z_{\tilde{\nu}}^{4j*}\right)P_R \bigg].
\end{eqnarray*}
		
\item {$(W_L^- - \chi^0 - \chi^-)$}\\
$\rightarrow$ $V_{\left(W_L^--\chi^0-\chi^-\right)} = \frac{ie}{s_W}\gamma^{\mu}\left[\left(Z_{\tilde{\nu}}^{2i}Z_{\tilde{\chi}^+}^{1j*} - \frac{1}{\sqrt{2}}Z_{\tilde{\nu}}^{4i}Z_{\tilde{\chi}^+}^{2j*}\right)P_L + \left(Z_{\tilde{\nu}}^{2i*}Z_{\tilde{\chi}^-}^{1j} + \frac{1}{\sqrt{2}}Z_{\tilde{\nu}}^{3i*}Z_{\tilde{\chi}^-}^{2j}\right)P_R\right]$.
	
\item {$(e^- - \chi^0 - \tilde{e})$}\\
\begin{eqnarray*}
\hspace{-35mm}	\rightarrow V_{\left(e^- - \chi^0 - \tilde{e}\right)}  &=& i\bigg[\left(\frac{-e}{\sqrt{2}s_Wc_W}(Z_{\tilde{\nu}}^{1j}s_W + Z_{\tilde{\nu}}^{2j}c_W) + Y_L^IZ_{\tilde{L}}^{i(I+3)}Z_{\tilde{\nu}}^{3j}\right)P_L  \nonumber \\
&&+ \left(\frac{-e\sqrt{2}}{c_W}Z_{\tilde{L}}^{i(I+3)}Z_{\tilde{\nu}}^{1j*} + Y_L^IZ_{\tilde{L}}^{Ii}Z_{\tilde{\nu}}^{3j*}\right)P_R\bigg].
\end{eqnarray*}
	
\item {$(W_L^- - \tilde{\nu} - \tilde{e})$}\\
$\rightarrow$ $V_{\left(W_L^- - \tilde{\nu} - \tilde{e}\right)} = \frac{-ie}{\sqrt{2}s_W}Z_{\tilde{\nu}}^{IJ}Z_{\tilde{L}}^{Ii}(p+k)^{\mu}$.
\end{enumerate}	
	

\end{document}